\documentclass[prb,floatfix,showpacs,superscriptaddress,twocolumn]{revtex4-1}
\usepackage{amsmath}
\usepackage{amssymb}
\usepackage{amsfonts}
\usepackage{graphicx}
\usepackage{dcolumn}
\usepackage{bm}
\usepackage{color}
\usepackage{chngcntr}

\DeclareGraphicsExtensions{.pdf,.png,.jpg,.eps}
\def\e{\mathop{\rm \mbox{{\Large e}}}\nolimits}

\begin{document}

\title{Electron-phonon deformation potential interaction in core-shell Ge-Si and Si-Ge nanowires}
\author{Dar\'{\i}o G. Santiago-P\'{e}rez}
\thanks{Corresponding author. Permanent address: U. Sancti Spiritus, Cuba.\\
e-mails: dariog@cbpf.br, dario@uniss.edu.cu}
\affiliation{Universidad de Sancti Spiritus ``Jos\'{e} Mart\'{\i} P\'{e}rez", Ave. de los M\'artires 360, CP 62100, Sancti Spiritus, Cuba}
\affiliation{CLAF - Centro Latino-Americano de F\'{\i}sica, Avenida Venceslau Braz, 71, Fundos, 22290-140, Rio de Janeiro, RJ, Brasil}
\thanks{Corresponding author. Permanent address: U. Sancti Spiritus, Cuba.\\
e-mails: dariog@cbpf.br, dario@uniss.edu.cu}
\author{C. Trallero-Giner }
\affiliation{Department of Theoretical Physics, Havana University, Havana 10400, Cuba}
\affiliation{Departamento de F\'{\i}sica, Universidad Federal de S\~{a}o Carlos, 13 565-905, S\~{a}o Carlos, Brazil}
\author{R. P\'{e}rez-\'Alvarez}
\affiliation{Universidad Aut\'{o}noma del Estado de Morelos, Ave. Universidad 1001, CP 62209, Cuernavaca, Morelos, M\'{e}xico}
\author{Leonor Chico}
\affiliation{Instituto de Ciencia de Materiales de Madrid (ICMM), Consejo Superior de Investigaciones Cient\'{\i}ficas
(CSIC), C/ Sor Juana In\'es de la Cruz 3, 28049 Madrid, Spain}
\author{G. E. Marques}
\affiliation{Departamento de F\'{\i}sica, Universidad Federal de S\~{a}o Carlos, 13 565-905, S\~{a}o Carlos, Brazil}
\date{\today }

\begin{abstract}
We settle a general expression for the Hamiltonian of the electron-phonon deformation potential (DP) interaction in the case of non-polar core-shell cylindrical nanowires (NWs). On the basis of long range phenomenological continuum model for the optical modes and by taking into account the bulk phonon dispersions, we study the size dependence and strain-induced shift of the electron-phonon coupling strengths for Ge-Si and Si-Ge NWs. We derive analytically the DP electron-phonon Hamiltonian and report some numerical results for the frequency core modes and vibrational amplitudes. Our approach allows for the unambiguous identification of the strain and confinement effects. We explore the dependence of mode frequencies and hole-DP scattering rates on the structural parameters of these core-shell structures, which constitute a basic tool for the characterization and device applications of these novel nanosystems.
\end{abstract}

\maketitle

\section{\label{Introduction}Introduction}

Semiconductor nanowires are at the focus of intense research due to their potential design of nanoscale devices, with applications in electronics, photonics, and nanosensors; besides, they constitute unique systems to explore novel low-dimensional phenomena, with great basic interest.~\cite{WuLieber2006,Hayden2008,AdvmatDasgupta2014} The experimental fabrication of core-shell nanowires has expanded the possibilities for tailoring the physical properties of these structures. Systems composed by Si, Ge and their solid solutions, are between the most studied and emerging as natural choices for integration with Si-based electronics.
The successful synthesis of Si-Ge core-shell nanowires~\cite{Lauhon2002} and the variety of applications foreseen for these materials has boosted the interest of many researchers.~\cite{SinghSST2012, PhysRevB.86.045311, PhysRevB.86.085318, singh:124305, MartinezGutierrez201386,
PhysRevB.84.085442, Liu20083042, PhysRevB.80.155432, PhysRevB.71.155318, PhysRevB.81.174303} Distinct physical properties, such as the separation of electron and hole carriers or the dramatic reduction of the thermal conductivity, are attained in Ge-Si core-shell NWs. Furthermore, with this cylindrical geometry it is possible to achieve much higher strains between the two materials without losing crystalline coherence,~\cite{Trammell2008} which can be of interest to modify the carrier mobility and effective masses in these nanostructures. However, there are limits to the wire diameters that can be grown without yielding defects, such as dislocations at the interface and shell corrugation, in order to relax the stress.~\cite{Goldthorpe2008} The crystalline orientation of the nanowire is another parameter to be considered.
In fact, the study of acoustic phonons in strained Si-Ge nanowires has been recently addressed  by means of a phenomenological continuum model.~\cite{Loss2014}
Strain may affect the lifetimes of spin qubits and it has important consequences in the electronic and optical properties.~\cite{Loss2012}

In order to characterize core-shell nanowires, Raman spectroscopy, a nondestructive technique, as well as infrared polarizability (IRP) are widely used to provide information on the phonon response region, the differences between various confined optical vibrations, their angular momentum, dependence size and structural effects and type of semiconductors involved in a structure. In order to elucidate the Raman selection rules, phonon scattering rates, confinement and strain effects in these systems, the knowledge of the electron-phonon Hamiltonian (EPH) as well as the optical modes of the nanostructure are necessary. By employing a continuum model, we aim at a description of EPH and the dependence of the optical modes with wire radii and phonon symmetry for non-polar materials.

It is well-known (see Ref. \onlinecite{PhysRevB.78.081304} and references therein) that in III-V and II-VI semiconductor nanostructures,  the Fr\"{o}hlich-like long range electrostatic potential is the most relevant interaction. In Si-Ge and Ge-Si, being non-polar materials, the electrostatic contribution due to the anion-cation atomic vibrations is absent. Consequently, the dominant contribution to the EPH is the mechanical deformation potential.~\cite{bir1974symmetry} In this sense, for a reliable description of the interaction between non-polar vibrations and electronic quasi-particles,
it is necessary the knowledge of phonon displacement vectors and their spatial symmetries. For the particular case of the electron-optical phonon Hamiltonian, these characteristics determine other physical properties, such as hole scattering, transport, Raman efficiency, IRP and Raman selection rules. Hence, an straightforward explicit expression for the EPH, as well as the understanding of its physical relevance, represent a central issue for the investigation of these novel structures.

In this work we study the optical modes and the corresponding optical deformation potential electron-phonon Hamiltonian of core-shell nanowires based on Si and Ge. We address the frequencies, phonon amplitudes, and symmetry dependence on core modes with respect to the relative dimensions of the system, i.e., core radius, shell thickness, ratio between core and shell radii, and the subsequent stress which builds up at the core-shell interface. We analyze the coupling between modes and the dispersion relations for these structures. We focus on core modes, for which the strain is homogeneous, in contrast to shell modes, which present a radial dependence on strain and, thus, making more difficult to distinguish between the contributions of strain and confinement for characterization purposes.~\cite{Trammell2008,Menendez} To this end, we employ a continuum approach, as has been done for other systems,~\cite{CubaLibro} including core-shell nanowires of polar semiconductors.~\cite{santiago-perez:084322,SantiagoPerez2014151} As in the polar case, both core and shell components develop strain due to the different lattice constant between the two materials. We include this effect in our model, so that frequencies at the center of the Brillouin zone of the bulk material are shifted with respect to the unstrained case. Thus, a macroscopic treatment of the phonon confinement frequencies and their spatial eigensolutions becomes a powerful tool to tackle the electron-phonon Hamiltonian in cylindrical core-shell NWs.

This work is organized as follows: Sec.~\ref{Electron-Optical} addresses the main formalism used to obtain the optical deformation potential Hamiltonian interaction for cylindrical nanowires. Furthermore, we provide an explicit analytical equation for the hole scattering matrix elements in terms of the $4\times4$ Luttinger Hamiltonian, deformation potential tensor and phonon field displacement. Sec.~\ref{model} presents the details of the phenomenological model. A brief review is given in Sec.~\ref{eqsandsols}, showing the equations of motion and the explicit form of the basis set for the solutions. Sec.~\ref{FrequencyShift} details the inclusion of strain effects on the vibrational frequencies of the corresponding bulk materials. In Sec.~\ref{Modos} we present analytical results for particular cases of the phonon dispersions relations with higher symmetry, which allows us to evaluate the shift due to confinement effects and strain, as well as the coupling between vibrational modes. Additionally, numerical results for Ge-Si and Si-Ge nanowires are shown. Section~\ref{Matrix-Element} is devoted to a direct evaluation of hole scattering rates due to deformation potential interaction Hamiltonian for the main phonon modes in the NWs. Finally, we draw our conclusions in Sec.~\ref{Conclusions}.

\section{\label{Electron-Optical}Electron-Optical phonon interaction in
core-shell nanowires}

In non-polar semiconductors, the deformation potential is a short-range interaction.~\cite{bir1974symmetry} Thus, in
the framework of the Born-Oppenheimer linear approximation,  the electron-phonon interaction can be written as
\begin{equation}
H_{\rm e-ph}=\vec{u}\cdot \frac{\partial H}{\partial \vec{u}}\;.
\label{Hep}
\end{equation}
Here, $\vec{u}$ is the phonon field displacement and $\partial H/\partial \vec{u}$ takes into account the perturbation of the electronic Hamiltonian by the optical phonon modes. From Eq.~(\ref{Hep}), matrix elements are formed between Bloch functions;
they depend on the phonon propagation and the crystal symmetry. For the diamond structure, the degenerate valence bands present $\Gamma_{25^{\prime }}$ symmetry at the $\Gamma-$point of the Brillouin zone.~\footnote{Due to symmetry reasons, the contribution of the conduction band at the $\Gamma$-point is zero.} The valence-band edge wavefunctions are given by
\begin{eqnarray}
|v_{-\frac{3}{2}}\rangle &=&\frac{i}{\sqrt{2}}|(x-iy)\rangle |\downarrow
\rangle \;,  \notag  \label{valence_band_funtions} \\
|v_{-\frac{1}{2}}\rangle &=&\frac{1}{\sqrt{6}}|(x-iy)\rangle |\uparrow
\rangle +\sqrt{\frac{2}{3}}|z\rangle |\downarrow \rangle \;,  \notag \\
|v_{\frac{1}{2}}\rangle &=&\frac{i}{\sqrt{6}}|(x+iy)\rangle |\downarrow
\rangle -i\sqrt{\frac{2}{3}}|z\rangle |\uparrow \rangle \;,  \notag \\
|v_{\frac{3}{2}}\rangle &=&\frac{1}{\sqrt{2}}|(x+iy)\rangle |\uparrow
\rangle \;,
\end{eqnarray}
where  $j=-\frac{3}{2},-\frac{1}{2},\frac{1}{2},\frac{3}{2}$ is the angular momentum quantum number,
and $|\uparrow \rangle $ ($|\downarrow \rangle $) denotes the spin parallel (antiparallel) to the growth direction $z$. In consequence, the deformation potential $\vec{\vec{D}}=\partial H/\partial \vec{u}$ can be characterized by the matrix elements between valence-band-edge wavefunctions $|v_{j}\rangle $. Under the symmetry operations of the representation
$\Gamma_{25^{\prime}}$, the only non-zero elements of the deformation potential tensor $\vec{\vec{D}}$ are $\langle y|D_{x}|z\rangle$, $\langle z|D_{y}|y\rangle $, $\langle y|D_{z}|x\rangle$ and equivalents.~\cite{Cardona} Hence, in matricial form, the $\vec{\vec{D}}$ components in cylindrical coordinates can be expressed as follows:
\begin{eqnarray}
D_{\hat{e}_{r}} &=&\frac{du_{0}}{a_{0}}\left(
\begin{array}{cccc}
0 & -\mathop{\rm \mbox{{\Large e}}}\nolimits^{i\theta } & 0 & 0 \\
-\mathop{\rm \mbox{{\Large e}}}\nolimits^{-i\theta } & 0 & 0 & 0 \\
0 & 0 & 0 & \mathop{\rm \mbox{{\Large e}}}\nolimits^{i\theta } \\
0 & 0 & \mathop{\rm \mbox{{\Large e}}}\nolimits^{-i\theta } & 0 \\
&  &  &
\end{array}
\right) \;,
\label{Matrix_Dr} \\
&&  \notag
\end{eqnarray}
\begin{eqnarray}
D_{\hat{e}_{\theta }} &=&\frac{idu_{0}}{a_{0}}\left(
\begin{array}{cccc}
0 & -\mathop{\rm \mbox{{\Large e}}}\nolimits^{i\theta } & 0 & 0 \\
\mathop{\rm \mbox{{\Large e}}}\nolimits^{-i\theta } & 0 & 0 & 0 \\
0 & 0 & 0 & \mathop{\rm \mbox{{\Large e}}}\nolimits^{i\theta } \\
0 & 0 & -\mathop{\rm \mbox{{\Large e}}}\nolimits^{-i\theta } & 0 \\
&  &  &
\end{array}
\right) \;,
\label{Matrix_Dcita} \\
&&  \notag
\end{eqnarray}
and
\begin{equation}
D_{\hat{e}_{z}}=\frac{idu_{0}}{a_{0}}\left(
\begin{array}{cccc}
0 & 0 & -1 & 0 \\
0 & 0 & 0 & -1 \\
1 & 0 & 0 & 0 \\
0 & 1 & 0 & 0 \\
&  &  &
\end{array}
\right) \;,
\label{Matrix_Dz}
\end{equation}

\noindent with $d$ being the optical DP constant as defined by Bir and Pikus,~\cite{bir1974symmetry} $a_{0}$ the lattice constant, $u_{0}=(\hbar V_{c}/VM\omega_{0})^{\frac{1}{2}}$ the unit of phonon displacement, $V_{c}$ the volume of the primitive cell, $M$ the atomic mass, $V$ the volume of the nanowire and $\omega_{0}$ the optical bulk phonon frequency at $\Gamma$-point.

The Hamiltonian for the electron-phonon interaction in the occupation number representation can be expressed as~\cite{Madel}

\begin{equation}
H_{\rm el-ph}=\sum_{\alpha ,\beta ,j,k_{z}}M_{\beta ,\alpha
}^{(j)}(a_{j}^{\dagger }(k_{z})+a_{j}(-k_{z}))c_{\beta }^{\dagger }c_{\alpha
}\;,
\label{Hel-ph}
\end{equation}

\noindent where $a_{j}(k_{z})^{\dagger }$($a_{j}(-k_{z})$) and $c_{\beta }^{\dagger }$($c_{\alpha }$) denote the phonon and electron creation (annihilation) operators in the branch $j$ with wavevector $k_{z}(-k_{z})$ and state $\beta $($\alpha $), respectively. In Eq.~(\ref{Hel-ph}), $M_{\beta ,\alpha}^{(j)}$ represents the amplitude probability of scattering between the
electronic states $\alpha \rightarrow \beta $ due to the interaction with an optical phonon with displacement $\vec{u}^{(j)}$. This probability amplitude is given by

\begin{equation}
M_{\beta ,\alpha }^{(j)}=\frac{1}{\sqrt{N_{j}}}\langle \beta |\vec{u}
^{(j)}\cdot \vec{\vec{D}}|\alpha \rangle \;,
\label{M_Matrixelem}
\end{equation}

\noindent where $N_{j}=\Vert \vec{u}^{(j)}\Vert $ is a normalization constant.

In the framework of the envelope function approximation for the $4\times 4$ Luttinger Hamiltonian~\cite{PhysRev.97.869} in the axial approximation,  and taking into account stress effects due to lattice mismatch, the fourfold wavefunction of the $\Gamma _{25^{\prime}}$ valence band states  can be expressed as

\begin{equation}
\langle \vec{r}|\alpha \rangle =\left(
\begin{array}{c}
F_{\nu }(r)|v_{\frac{3}{2}}\rangle \\
F_{\nu +1}(r)\mathop{\rm \mbox{{\Large e}}}\nolimits^{i\theta }|v_{\frac{1}{2}}\rangle \\
F_{\nu +2}(r)\mathop{\rm \mbox{{\Large e}}}\nolimits^{2i\theta }|v_{-\frac{1}{2}}\rangle \\
F_{\nu +3}(r)\mathop{\rm \mbox{{\Large e}}}\nolimits^{3i\theta }|v_{-\frac{3}{2}}\rangle \\
\end{array}
\right) \mathop{\rm \mbox{{\Large e}}}\nolimits^{i(k_{e}z+\nu \theta )}\;.
\label{function_rLunti}
\end{equation}

\noindent Here, $F_{\nu }(r)$ is the Bessel function $J_{\nu }$ for $r < a$ and a linear combination of Bessel and Neumann functions~\cite{Abramowitz} for $a < r < b$. Thus, the scattering matrix elements (\ref{M_Matrixelem}) can be cast as
\begin{widetext}
\begin{eqnarray}
\label{M}
M_{\beta,\alpha}^{(j)}=\frac{1}{\sqrt{N_{j}}}\left\langle\left(
                                \begin{array}{c}
                                  F_{\nu^{\prime}}(r)|v_{\frac{3}{2}}\rangle \\
                                  F_{\nu^{\prime}+1}(r)\e^{i\theta}|v_{\frac{1}{2}}\rangle \\
                                  F_{\nu^{\prime}+2}(r)\e^{2i\theta}|v_{-\frac{1}{2}}\rangle \\
                                  F_{\nu^{\prime}+3}(r)\e^{3i\theta}|v_{-\frac{3}{2}}\rangle \\
                                \end{array}
                              \right)\right|\vec{u}^{(j)}\cdot \vec{\vec{D}}\e^{i(\nu-\nu^{\prime})\theta}\left|\left(
                                \begin{array}{c}
                                  F_{\nu}(r)|v_{\frac{3}{2}}\rangle \\
                                  F_{\nu+1}(r)\e^{i\theta}|v_{\frac{1}{2}}\rangle \\
                                  F_{\nu+2}(r)\e^{2i\theta}|v_{-\frac{1}{2}}\rangle \\
                                  F_{\nu+3}(r)\e^{3i\theta}|v_{-\frac{3}{2}}\rangle \\
                                \end{array}
                              \right)\right\rangle\delta_{k_{e}^{\prime},k_{e}\pm k_{z}}\;,
\end{eqnarray}
\end{widetext} 

\noindent where the momentum conservation along the $z$-direction is written explicitly. The influence of the geometric factors, as well as the strain and bulk parameters on the matrix elements~(\ref{M}), are embedded in the phonon dispersion relations and the corresponding displacement vectors.

\section{\label{model} Phenomenological continuum approach in cylindrical geometry}

In order to derive a comprehensive expression for the electron-phonon DP matrix elements (\ref{M}), it
is required to discuss the phonon dispersion relations as a
function of radii $a$ and $b$, wavevector $k_z$, and influence of the strain effects across the core-shell surface, as well as the spatial symmetry properties of the phonon displacement vector. In the following, we study the confined phonon frequencies, the mixing of phonon modes as a consequence of the cylindrical spatial geometry and their corresponding displacement vector, based on a unified macroscopic continuum theory where the medium properties are considered to be piecewise.~\cite{Comas1993a,0953-8984-7-9-006}

\subsection{\label{eqsandsols} Equations of motion and basis for the solutions}

We consider infinite cylindrical core-shell nanowires with core radius $a$ and shell radius $b$, so that the shell thickness is given by $b-a$. We choose the axis of the wire along the $z$-direction of the cylindrical coordinates $(r,\theta,z)$. Although the continuum approach employed in this work has been reported elsewhere,~\cite{CubaLibro,PhysRevB.69.035419,SantiagoPerez2014151} for the sake of completeness and further applications focusing on the electron-phonon DP Hamiltonian, we briefly recall the main features of the model, particularizing for non-polar media and cylindrical core-shell geometry. Considering a harmonic time-dependence for the oscillations, the equations of motion for the optical modes in a isotropic non-polar media is given by~\footnote{This equation is straightforward derived from the hydrodynamic phenomenological model for cubic polar semiconductors described in Refs. ~\onlinecite{CubaLibro,PhysRevB.37.4583} considering that the polarization and electric field associated with vibrations are zero. Note that since the core and shell bulk materials are non-polar, $\omega _{TO}=\omega_{LO}=\omega_{0}$.}
\begin{equation}
(\omega ^{2}-\omega _{0}^{2})\vec{u}=\beta _{L}^{2}\nabla (\nabla \cdot \vec{u})-\beta _{T}^{2}\nabla \times \nabla \times \vec{u}\;.
\label{ecuacion de movimiento2}
\end{equation}

\noindent In these expressions, $\beta _{L}$, $\beta _{T}$ describe the quadratic dispersions of the LO- and TO-bulk phonon branches of the optical modes in the long-wave limit, respectively. Applying the Helmholtz's method of potentials,~\cite{Morse,PhysRevB.69.035419,SantiagoPerez2014151} one can find a general basis of solutions for the problem, namely
\begin{eqnarray}
\vec{u}_{T1} &=&\left(
\begin{array}{c}
\frac{ik_{z}}{q_{T}}f_{n}^{\text{ }\prime }(q_{T}r) \\
-\frac{nk_{z}}{q_{T}}\frac{1}{q_{T}r}f_{n}(q_{T}r)\nonumber \\
f_{n}(q_{T}r) \\
\end{array}
\right) e^{i(n\theta +k_{z}z)}\;,  \notag \\
\vec{u}_{T2} &=&\left(
\begin{array}{c}
\frac{in}{q_{T}r}f_{n}(q_{T}r) \\
-f_{n}^{\text{ }\prime }(q_{T}r) \\
0 \\
\end{array}
\right) e^{i(n\theta +k_{z}z)}\;,  \notag \\
\vec{u}_{L} &=&\left(
\begin{array}{c}
f_{n}^{\text{ }\prime }(q_{L}r) \\
\frac{in}{q_{L}r}f_{n}(q_{L}r) \\
\frac{ik_{z}}{q_{L}}f_{n}(q_{L}r) \\
\end{array}
\right) e^{i(n\theta +k_{z}z)}\;,
\label{basis}
\end{eqnarray}
\noindent where the vector components are in cylindrical coordinates, $(u_{r}, u_{\theta }, u_{z})$; the prime denotes the derivative with respect to the argument; $n$ is an integer label related to the angular dependence of the modes; $k_{z}$ the continuum wavevector along the cylinder axis, and the wavevectors $q_{L,T}$ are given by

\begin{equation}
\label{qlt}
q_{L,T}^{2}=\frac{\omega _{0}^{2}-\omega ^{2}}{\beta _{L,T}^{2}}-k_{z}^{2}\;.
\end{equation}

If $q_{L,T}^{2} > 0$ ($q_{L,T}^{2} < 0$) the function $f_{n}$ is an order-$n$ Bessel (modified Bessel) function of the first or second kind, i.e., Bessel $J_{n}$ or Neumann $N_{n}$ (Infield $I_{n}$ or MacDonald $K_{n}$). It is straightforward to check that the longitudinal solution verifies $\nabla \times \vec{u}_{L}=\vec{0}$, whereas the transverse solutions satisfy $\nabla \cdot \vec{u}_{T1}=\nabla \cdot \vec{u}_{T2}=0$, as it should be.
Particular cases of this basis have been used to study phonon modes in non-polar nanotubes~\cite{PhysRevB.69.035419,CPC2006} and in solid nanowires with only one material at $k_{z}=0.$~\cite{Comas1993a,0953-8984-7-9-006}

In cylindrical geometry, neither the amplitudes $\vec{u}_{T1}$, $\vec{u}_{T2}$ nor $\vec{u}_{L}$ represent independent solutions for the phonon modes of the core-shell nanostructures. Nevertheless, the explicit form of the basis (\ref{basis}) allows us to elucidate the uncoupled modes and their polarization for special symmetries, such as $n=0$ or $k_z=0$.

A direct evaluation of Eq.~(\ref{M}) leads to search for the general solution of the problem. This can be written as a linear combination of the basis vectors (\ref{basis}), whose coefficients are determined by imposing the appropriate boundary conditions. If the bulk optical frequencies of core and shell materials are very different, it is a valid assumption that states are completely confined in the core or in the shell regions. This assumption is completely fulfilled for Si and Ge, whose characteristic optical phonon frequencies are 521 and 301 cm$^{-1}$ respectively.~\cite{Madelung}
In addition,
we will assume a large separation between the optical branches of shell and the host material. Thus, the amplitude of the oscillations should be zero at the surfaces $S$ ($r=a$ and $r=b$), i.e., $\vec{u}|_S=0$.

\subsection{\label{FrequencyShift} Strain-induced shift of bulk modes}

Core-shell silicon and germanium NWs should present large strain fields due to the lattice mismatch at the interface. This effect has been measured by Raman spectroscopy,~\cite{SinghSST2012, singh:124305, Menendez} as well as the strain-induced frequency shift as a
function of core radius and shell thickness.~\cite{PhysRevB.86.045311} The frequency shift can be estimated by solving the secular equation~\cite{Anastassakis}

\begin{equation}
\left\vert
\begin{array}{ccc}
p\varepsilon _{11}+q\widetilde{\varepsilon _{11}}-\lambda & 2t\varepsilon_{12} & 2t\varepsilon _{13} \\
2t\varepsilon _{21} & p\varepsilon _{22}+q\widetilde{\varepsilon _{22}}-\lambda & 2t\varepsilon _{23} \\
2t\varepsilon _{31} & 2t\varepsilon _{32} & p\varepsilon _{33}+q\widetilde{\varepsilon _{33}}-\lambda \\
&  &
\end{array}
\right\vert =0\;,
\label{Shift}
\end{equation}

\noindent where $p$, $q$ and $t$ are the phonon deformation potential values, $\varepsilon _{ij}$$(i,j=1,2,3)$ the strain components in cartesian
coordinates, $\widetilde{\varepsilon _{ii}}=tr\{\varepsilon \}-\varepsilon _{ii}$, $tr\{\varepsilon \}$ is the trace of the stress tensor, and
$\lambda =\omega ^{2}-\omega _{0}^{2}$ is the strain-induced frequency shift. In the present work we will deal with nanowires grown along the [011] direction.
A detailed procedure for the evaluation of the shift $\lambda$ in
the above-mentioned crystallographic direction and analytical expressions for $\varepsilon _{ij}^{\mathrm{core}}$ and $\varepsilon _{ij}^{\mathrm{shell}}$ are given in Refs.~\onlinecite{Menendez,phononsonly}. Here we
present the corresponding solutions,

\begin{eqnarray}
\lambda _{L}&=&(\frac{3}{4}p+\frac{5}{4}q+\frac{1}{2}
t)\varepsilon _{rr}^{\rm core}+(\frac{1}{4}p+\frac{3}{4}q-\frac{1}{2}
t)\varepsilon _{zz}^{\mathrm{core}}\;,
\notag
\label{corrimiento-core011}
\\
\lambda _{T1}&=&(\frac{1}{2}p+\frac{3}{2}q-t)\varepsilon
_{rr}^{\rm core}+(\frac{1}{2}p+\frac{1}{2}q+t)\varepsilon _{zz}^{\mathrm{core}}\;,
\notag \\
\lambda _{T2}&=&(\frac{3}{4}p+\frac{5}{4}q+\frac{1}{2}
t)\varepsilon _{rr}^{\rm core}+(\frac{1}{4}p+\frac{3}{4}q-\frac{1}{2}
t)\varepsilon _{zz}^{\mathrm{core}}\;.
\notag \\
\end{eqnarray}

Notice that the frequency shift in the core only depends on the ratio $\gamma$, and not on the particular values of the core and shell radii. However, for the shell,
$\varepsilon_{rr}^{\rm shell}$ and $\varepsilon_{\theta\theta}^{\rm shell}$ depend on the coordinate $r$. For this reason, $\lambda_{1}^{\rm shell}$ and $\lambda_{2}^{\rm shell}$ are non-trivial functions of $r$ and $\theta$. As in this work we focus on core modes, it is sufficient with the expressions (\ref{corrimiento-core011}) shown above.

Studies by Raman spectroscopy prove that strain is partially relaxed, at least for the core diameters experimentally obtained to this date.
In order to model this effect, Singh \textit{et al.} ~\cite{SinghSST2012} introduced an axial relaxation parameter $\rho$ in the misfit factor, $\varepsilon_{m}=\varepsilon_{zz}^{\rm core}-\varepsilon_{zz}^{\rm shell}$. In the framework of this heuristic approach, the misfit strain
is rewritten as $\varepsilon_{m}\rightarrow\varepsilon_{m}(1-\rho)$. This parameter varies between 0 and 1, so that when $\rho=0$, the system is fully strained. Since all the experimental information available to nowadays deals with nanowires with partially relaxed strain, we take for our numerical evaluations a relaxation parameter $\rho =0.5$, avoiding the unrealistic overestimation of the strain. The results for fully strained NWs are very similar, save for the larger shift due to strain effects.

Once the phonon bulk frequencies are corrected including strain through the replacement $\omega_{0}^{2}\rightarrow \omega_{0}^{2}+\lambda_{i}$ $(i=L,T1,T2)$ in the corresponding expressions (\ref{qlt}), we calculate the phonon dispersion relations using Eq.~(\ref{disper1}). In the following, we address some numerical results focusing on the higher symmetry modes.

\section{\label{Modos}Dispersion relations for core-shell nanowires}

We study the core modes in Ge-Si and Si-Ge systems and, in particular, we analyze the coupling for different values of $n$ and $k_{z}$, as well as the frequency shift due to confinement as a function of the core and shell radii $a$, $b$, and the wavevector $k_{z}$. Taking a linear combination of the basis functions (\ref{basis}) and applying the boundary condition $\vec{u}|_{r=a}=0$ , the general dispersion relations for core phonons are obtained by solving the transcendental equation

\begin{multline}
J_{n}(\mu_{T1})\left[J_{n}^{\text{ }\prime }(\mu_{L})J_{n}^{\text{ }\prime }(\mu_{T2})-\frac{n^{2}}{\mu_{L}\mu_{T2}}J_{n}(\mu_{L})J_{n}(\mu_{T2})\right] \\
=\frac{\widetilde{k}_{z}^{2}}{\mu_{L}\mu_{T1}}J_{n}(\mu_{L})\left[J_{n}^{\text{ }\prime }(\mu_{T1})J_{n}^{\text{ }\prime }(\mu_{T2})\right. \\
\left.-\frac{n^{2}}{\mu_{T1}\mu_{T2}}J_{n}(\mu_{T1})J_{n}(\mu_{T2})\right] \text{ },
\label{disper1}
\end{multline}
where $\widetilde{k}_{z}=k_{z}a$ and $\mu_{i}^{2}=q_{i}^{2}a^{2}+\lambda_{i}(\gamma )a^{2}/\beta _{i}^{2}$, ($i=L, T1, T2$).

From the above equation immediately follows the following symmetry properties:
\\
\noindent (i) for $n=0$ and $k_{z}=0$, the triple degeneracy of the optical modes is broken, and we have three independent
subsets of confined modes for $L$, $T1$, and $T2$;
\\
\noindent (ii) for $n \neq 0$ and $k_{z}=0$, the degeneracy is partially lifted:  $L$ and $T2$ modes are coupled, while $T1$ remains uncoupled;
\\
\noindent (iii) for $n=0$ with $k_{z} \neq 0$, the bulk degeneracy is also split into two subsets, one belonging to the independent transversal $T2$ phonon mode, and the
other corresponding to the coupled longitudinal and transverse $L-T1$ modes; and finally,
\\
\noindent (iv) for $n\neq 0$ and $ k_{z}\neq 0$ all the $L$, $T1$ and $T2$ phonon vector amplitudes are mixed.

These results, stemming from the peculiarities of the cylindrical geometry, have profound consequences on the $H_{\rm e-ph}$ Halmitonian~(\ref{Hep}). According to these symmetries, which are characterized by the azimuthal label $n$ and wavevector $k_z$,
four different physical situations can be distinguished in relation to the EPH, which will be of use to analyze subsequent calculations of the dispersion relations for core-shell Ge-Si and Si-Ge NWs.

Table ~\ref{Parametros} show the input parameters employed in the calculations.
In the following
calculations the values given in Tables~\ref{Parametros} are
assumed to be size-independent,
a hypothesis that should not be valid
for
very
small radii. Dimensionless quadratic curvature parameters for the transversal ($\beta_{T}^{2}$) and longitudinal ($\beta_{L}^{2}$) bulk optical phonon bands, along the [011] crystallographic direction employed in this work, are $6.33\times10^{-12}$, $11.53\times10^{-12}$  and $17.59\times10^{-12}$, $31.95\times10^{-12}$ for Ge and Si respectively. These values have been fitted to the neutron dispersion data collected in Ref.~\onlinecite{Hummer2009}, originally reported in Refs.~\onlinecite{Nilsson1972,Kulda1994} (Si) and \onlinecite{Nilsson1971} (Ge). As we know along the [011] crystallographic direction, the transversal optical phonons are non-degenerate and showing different $\beta_{T1}$ and $\beta_{T2}$ curvatures. For Si and Ge bulk semiconductors these values are close. Thus, in our calculations and in the framework of the isotropic approximation, we have chosen for $\beta_{T}$ the average values of $\beta_{T1}$ and $\beta_{T2}$ fitted by neutron scattering.

\begin{center}
\begin{table}[htb]
\caption{\label{Parametros} Bulk parameters for Ge and Si with diamond structure. $\omega_{0}$ is given in cm$^{-1}$, the Young's modulus $E$ in 10$^{12}$ dyn/cm$^{2}$ and the lattice constant $a_0$ in nm. (a) Ref.~\onlinecite{Madelung}; (b) Ref.~\onlinecite{Anastassakis}; (c) Ref.~\onlinecite{Adachi}.\\}
\begin{tabular}{ccccccccc}
  \hline
  \hline
$$ &  $$  & $\omega_{0}$  & $p/\omega_{0}^{2}$ & $q /\omega_{0}^{2}$  & $t /\omega_{0}^{2}$ &  $E$ & $\nu$ & $a_{0}$  \\
  \hline
  \hline
  Ge    & $$ & 301$^{a}$  & $-1.47^{b}$  &  $-1.93^{b}$ & $-1.11^{b}$   & 1.28$^{c}$ &   0.21$^{c}$ & 0.566$^{c}$  \\
  Si    & $$ & 521$^{a}$  & $-1.83^{b}$  &  $-2.33^{b}$ & $-0.71^{b}$  &  1.59$^{c}$ &   0.23$^{c}$ & 0.543$^{c}$  \\
  \hline
  \hline
\end{tabular}
\end{table}
\end{center}

\subsection{Modes with $n=0$ and $k_{z}=0$}

Firstly, we focus on the uncoupled modes with $n=0$ and $k_{z}=0$. By inspection of the basis for the solutions, it is clear that for this case all modes $L,$ $T1,$ and $T2$ are completely decoupled. Imposing the boundary condition of complete confinement, the frequencies of core modes are found to be
\begin{eqnarray}
\omega _{L}^{2} &=&\omega _{0}^{2}-\frac{\beta _{L}^{2}(\mu _{1}^{(m)})^{2}}{a^{2}}+\lambda _{L}(\gamma )\;,  \notag
\label{core-frecuencia} \\
\omega _{T1}^{2} &=&\omega _{0}^{2}-\frac{\beta _{T}^{2}(\mu _{0}^{(m)})^{2}}{a^{2}}+\lambda _{T1}(\gamma )\;, \\
\omega _{T2}^{2} &=&\omega _{0}^{2}-\frac{\beta _{T}^{2}(\mu _{1}^{(m)})^{2}}{a^{2}}+\lambda _{T2}(\gamma )\;,  \notag
\end{eqnarray}%
where $\mu _{i}^{(m)}$ $(i=0,1)$ are the roots of $J_{i}(\mu _{i}^{(m)})=0$, with $m=1, 2,...$.

The second term in the right hand side of Eqs.~(\ref{core-frecuencia}) gives the effect of confinement. Obviously, it is always negative, producing a downshift of the modes. The confinement term for these uncoupled modes varies with $1/a^2$. The third term is the effect of strain, $\lambda_{i}$, which depends on the ratio $\gamma$ and the crystallographic direction.

In the present case the $H_{\rm e-ph}$ is decoupled into three independent Hamiltonians, $H_{\rm e-ph}^{L}$, $H_{\rm e-ph}^{T1}$ and $H_{\rm e-ph}^{T2}$, characterizing the three orthogonal phonon displacements along the radial ($\hat{e}_{r}$), axial ($\hat{e}_{z}$) and azimuthal ($\hat{e}_{\theta }$) directions, respectively.

\begin{figure}[htb]
\begin{center}
\includegraphics[width=\columnwidth]{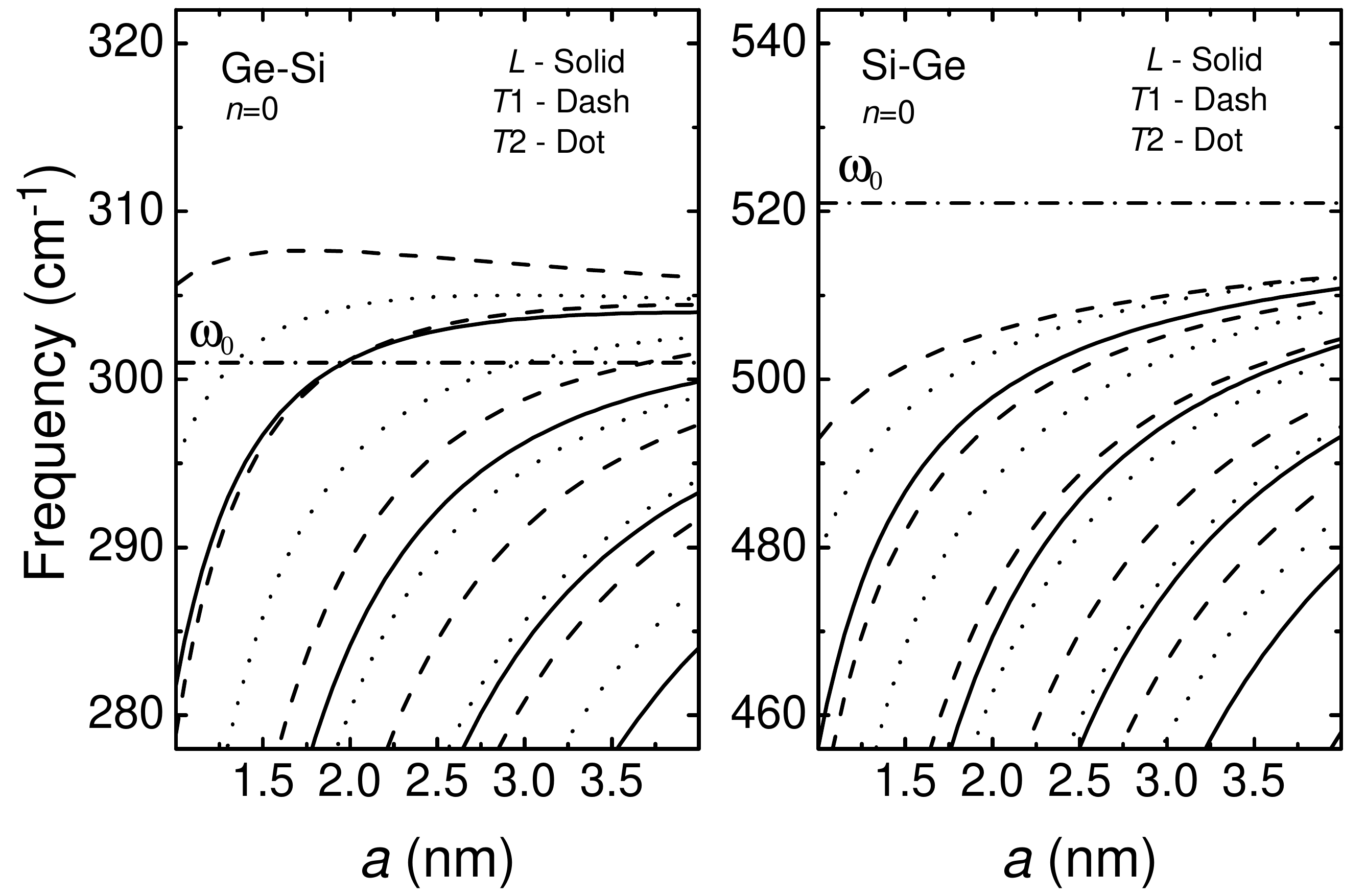}
\end{center}
\par
\caption{Frequencies of the core modes with $n=0$ and $k_{z}=0$ as a function of the core radius $a$ in a core-shell system grown in the [011] direction. Left panel: Ge-Si; right panel: Si-Ge, for fixed shell thickness $b-a=1$ nm. }
\label{NceroKzCeroRadio}
\end{figure}

Figure~\ref{NceroKzCeroRadio} shows the core modes as a function of the core radius $a$ in a core-shell system for fixed shell thickness. The left panel presents the Ge-Si case, and the right panel depicts results for the Si-Ge nanowire. Recall that the role of the shell is essential to obtain the shift of the core bulk frequency, as explained in Sec.~\ref{FrequencyShift} but, besides that, it does not play any role for the core modes, because of the boundary condition of complete confinement. There is an overall increase of the core mode frequencies in the left panel of Fig.~\ref{NceroKzCeroRadio}, in which Ge is the core material, while the modes are downshifted in the right panel of Fig.~\ref{NceroKzCeroRadio}, where Si is the core medium. This is related to the difference of lattice constants of Si and Ge; as it can be seen in Table~\ref{Parametros}, the lattice constant of Si is smaller than that of Ge, thus the strain always produces a redshift in the Si part of the wire, and a blueshift in the Ge part, no matter whether they constitute the core or the shell. The highest frequency mode of the Ge-core case (left panel) shows an increase of frequency for diminishing $a$ in a substantial radius range, due to the importance of strain for this mode.

Comparison of the results of Fig.~\ref{NceroKzCeroRadio} to the frequencies obtained for fixed shell/core ratio $\protect\gamma$ (not shown) allows us to conclude that for increasing values of $a$ and fixed shell thickness, the frequencies tend to the bulk core value, while for $\gamma$ fixed confinement effects disappear, leaving the strain as the main contribution. As in Fig.~\ref{NceroKzCeroRadio}, Ge core modes are blueshifted due to strain, whereas the Si modes are redshifted. The higher frequency mode of this latter panel also shows a blue shift for diminishing radius, which signals the prevalence of strain effects for this mode.

In a nanowire with fixed core radius, the frequency dependence is due to the strain, which varies with the shell radius via the ratio $\gamma$. As discussed above, the NW with Ge core will always shows an increasing blueshift of all modes with increasing strain, because of the smaller Si lattice constant. For the same reason, all modes of strained Si-core NWs are redshifted.

\subsection{Modes with $n\neq0$ and $k_{z}=0$}

In the case of modes without axial symmetry, i.e., $n \neq 0$, we find for $k_{z}=0$
that $L$ and $T2$ modes are coupled, while the $T1$ mode remains uncoupled. The dispersion relation for the latter is given by

\begin{eqnarray}
\label{T1desacoplado}
\omega^{2}_{T1}&=&\omega^{2}_{0}-\frac{\beta_{T}^{2}(\mu_{n}^{(m)})^{2}}{a^{2}}+\lambda_{T1}(\gamma) \;,
\end{eqnarray}

\noindent where $J_{n}(\mu_{n}^{(m)})=0$ with $m=1,2,...$ .

The coupled $L-T2$ modes fulfill the equation

\begin{eqnarray}  \label{LT2acoplados}
J_{n}^{\prime }(\mu_{L})J_{n}^{\prime }(\mu_{T2})-\frac{n^{2}}{\mu_{L}\mu_{T2}}{J}_{n}(\mu_{L}){J}_{n}(\mu_{T2})=0\;,
\end{eqnarray}

\noindent with

\begin{eqnarray}
\mu_{L}^{2}=(\omega^{2}_{0}+\lambda_{L}(\gamma) - \omega^{2}) \left(\frac{a}{\beta_{L}}\right)^{2}, \\
\mu_{T2}^{2}=(\omega^{2}_{0}+\lambda_{T2}(\gamma) - \omega^{2})\left(\frac{a}{\beta_{T}}\right)^{2}\;.
\end{eqnarray}

\begin{figure}[htb]
\begin{center}
\includegraphics[width=\columnwidth]{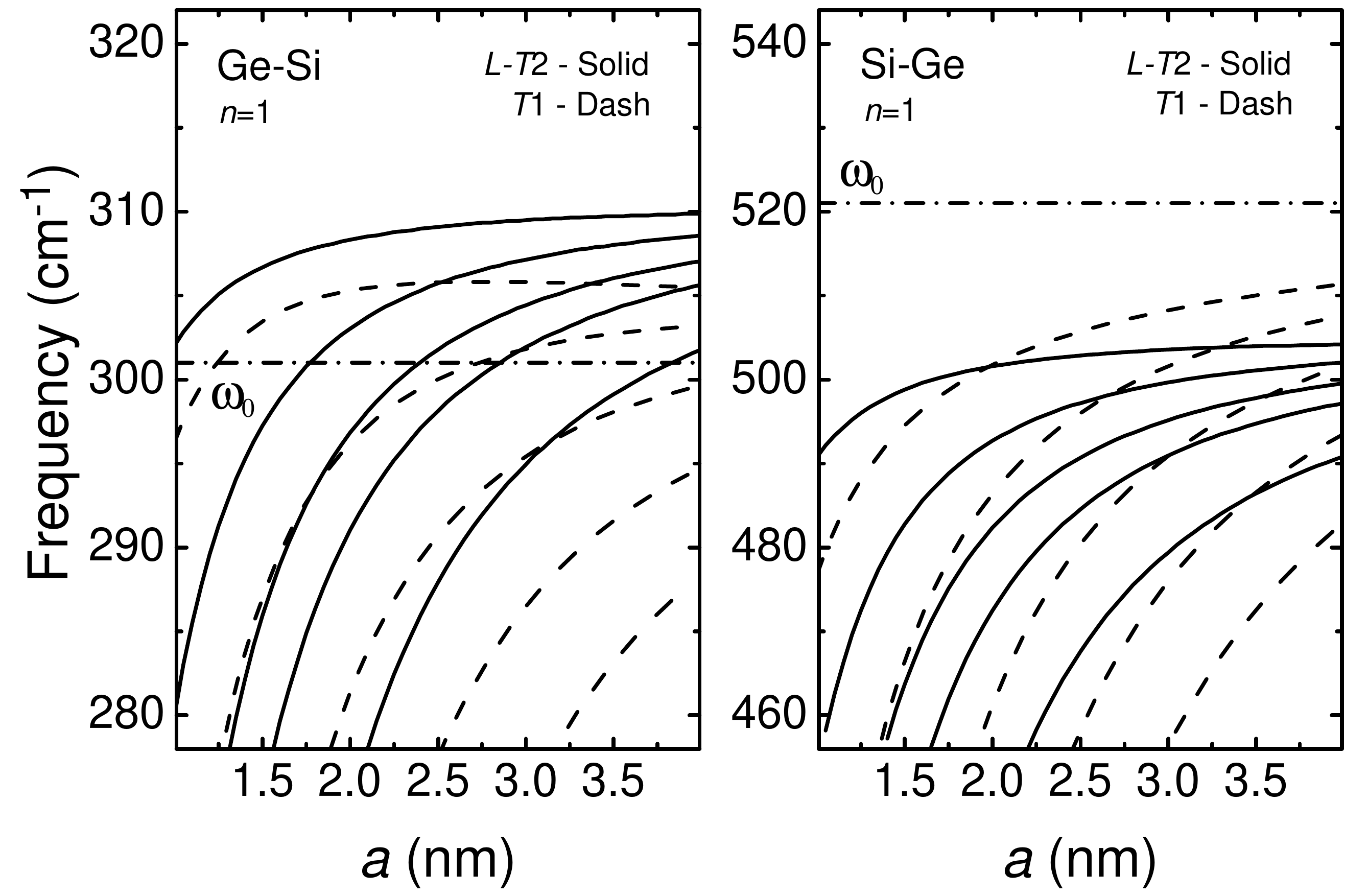}
\end{center}
\par
\caption{The same as Fig.~\ref{NceroKzCeroRadio} for $n=1$.}
\label{NunoKzCeroRadio}
\end{figure}

Here, we have only two independent blocks in the EPH $H_{\rm e-ph}$. One corresponds to $H_{\rm e-ph}^{T1}$ and the other to a mixture of $u_L$ and $u_{T2}$ amplitudes, with phonon  polarization vector on the ($\hat{e}_{r}$,$\hat{e}_{\theta }$) plane, which leads to $H_{\rm e-ph}^{L-T2}$.
Figure~\ref{NunoKzCeroRadio} shows the core modes with $n=1$ and $k_{z}=0$ as a function of the core radius for fixed shell thickness. Notice that the uncoupled $T1$ modes behave as for the $n=0$ case. The coupled $L-T2$ modes are closer in frequencies compared to the $n=0$ case. This behavior holds for varying core radius if the same shell/core ratio is maintained.

\subsection{Modes with $n=0$ and $k_{z}\neq0$}

Now we consider the dependence of the mode frequencies with the wavevector, $k_{z}\neq0$. We focus on $n=0$ modes. We obtain an uncoupled $T2$ mode and a coupled $L-T1$ mode. The uncoupled transverse mode is given by $J_{1}(\mu_{1}^{(m)})=0$, which leads to the dispersion relation

\begin{eqnarray}
\label{T2desacoplado}
\omega^{2}_{T2}&=&\omega^{2}_{0}-\frac{\beta_{T}^{2}(\mu_{1}^{(m)})^{2}}{a^{2}}+\lambda_{T2}(\gamma)-\beta_{T}^{2}k_{z}^{2} \;.
\end{eqnarray}

Equation~(\ref{T2desacoplado}) is just like the bulk dispersion relation, except for the shifts due to the spatial confinement
$(\beta_{T}\mu_{1}^{(m)})^{2}/a^{2}$ and the strain, $\lambda_{2}(\gamma)$. The coupled $L-T1$ modes are obtained from Eq.~(\ref{LT1acoplado}):

\begin{eqnarray}
\label{LT1acoplado}
J_{0}^{\prime }(\mu_{L})J_{0}(\mu_{T1})-\frac{\widetilde{k}_{z}^{2}}{\mu_{L}\mu_{T1}}J_{0}(\mu_{L})J_{0}^{\prime }(\mu_{T1})=0\;,
\end{eqnarray}
with
\begin{eqnarray}
\mu_{L}^{2}&=&(\omega^{2}_{0}+\lambda_{L}(\gamma) - \omega^{2})\left(\frac{a}{\beta_{L}}\right)^{2}-k_{z}^{2}a^{2}, \\
\mu_{T1}^{2}&=&(\omega^{2}_{0}+\lambda_{T1}(\gamma)-\omega^{2})\left(\frac{a}{\beta_{T}}\right)^{2}-k_{z}^{2}a^{2}\;.
\end{eqnarray}

If $k_z\neq0$, the axial symmetry is broken and for  $n=0$, the amplitudes $u_L$ and $u_{T1}$ are coupled, so we obtain the $H_{\rm e-ph}^{L-T1}$ which describes the electron interaction with phonons polarized on the ($\hat{e}_{r}$,$\hat{e}_{z}$) plane. Besides, we have a $H_{\rm e-ph}^{T2}$ term for the uncoupled $T2$ optical modes.

\section{\label{Matrix-Element}Electron-phonon scattering rate}

Notice that the influence of the geometric factors, as well as the strain and bulk parameters on  the electron-phonon
matrix elements (\ref{M}) are embedded in the phonon dispersion relations and the corresponding phonon displacement vector.
Hence, on the basis of the calculated frequencies and phonon amplitudes, explicit
expressions for the DP matrix elements (\ref{M_Matrixelem}) can be carried forward.
From Eq.~(\ref{M}) and the previous discussions, it becomes clear that the electron-phonon scattering rate
depends on the phonon polarization. Since we are in a cylindrical geometry,
it is not possible to decouple the phonon modes in a set of three independent polarizations. In the following subsections
we illustrate some cases of interest for the hole scattering
caused by the phonon polarization along the axial, radial and azimuthal directions.

\subsection{Phonon modes polarized along the growth direction}

For phonon modes polarized along the cylinder axis, we have to consider the $z$ component of the vector
amplitude $\vec{u}^{(j)}$. Thus, from the basis vectors shown in Eq.~(\ref{basis}) we have

\begin{multline}
u_{z}^{(\hat{e}_{z})}=U_{z}e^{in\theta }=\left( J_{n}(\mu_{T1}r/a)J_{n}^{\text{ }
}(\mu_{L})\right. \\
\left. -J_{n}(\mu_{L}r/a)J_{n}^{\text{ }}(\mu_{T1})\right) e^{in\theta
}/\sqrt{N_{z}}\text{ }.  \label{UZ}
\end{multline}%
Consequently, combining Eqs.~(\ref{Matrix_Dz}) and (\ref{function_rLunti}), the scattering amplitude (\ref{M}) can be
cast as
\begin{multline}
M_{\beta ,\alpha }^{(\hat{e}_{z})}=\left( \delta _{\nu ^{\prime },\nu +n+2}
\left[ -\langle \nu ^{\prime }|U_{z}|\nu +2\rangle -\right. \right. \\
\left. \langle \nu ^{\prime }+1|U_{z}|\nu +3\rangle \right] +\delta _{\nu
^{\prime },\nu +n-2}\left[ \langle \nu ^{\prime }+2|U_{z}|\nu \rangle \right.
\\
\left. \left. +\langle \nu ^{\prime }+3|U_{z}|\nu +1\rangle \right] \right)
\delta _{k_{e}^{\prime },k_{e}\pm k_{z}}\text{ }.  \label{Mz}
\end{multline}

This scattering rate is ruled by the combination of longitudinal
$L$ and transverse $T1$ amplitudes. In the particular case of $k_{z}=0$,
as it is required for example in infrared spectroscopy measurements,
the hole transition is assisted by a pure transversal $T1$ optical phonon.

\begin{figure}[tbh]
\begin{center}
\includegraphics[width=\columnwidth]{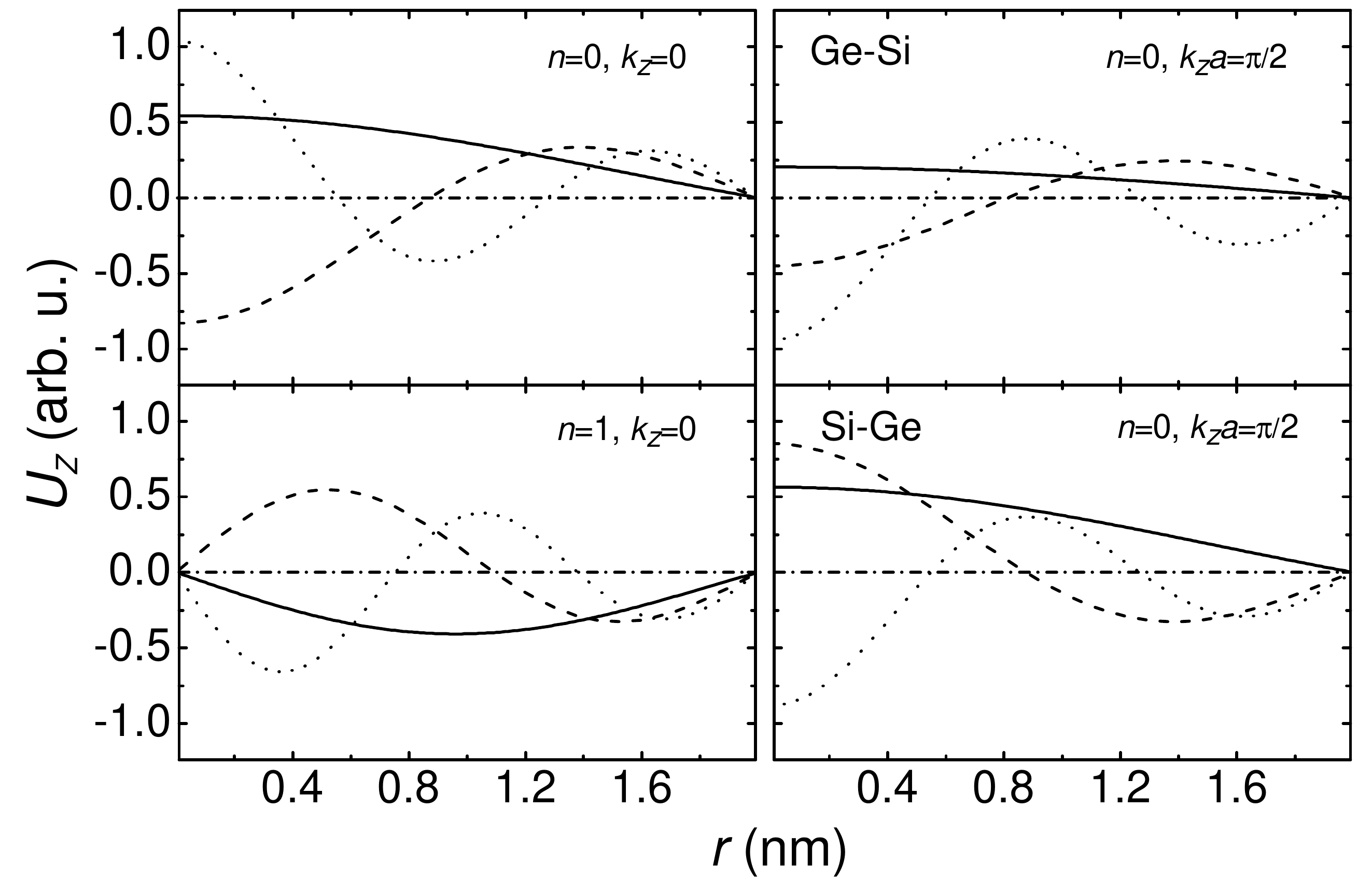}\\[0pt]
\end{center}
\caption{Core phonon amplitude $U_z$ for Ge-Si and Si-Ge core-shell NWs. Left panel: $n=0$, 1 and $k_{z}=0$; right panel: $n=0$ and $k_{z}a=\protect\pi/2$. In the calculation $a=2$ nm and $b=4$ nm.}
\label{Uz}
\end{figure}

Figure \ref{Uz} shows the contribution of the amplitude $u_z$ of the modes polarized along the core-shell growth direction to the EPH
$H_{\rm e-ph}^{(\hat{e}_{z})}$. The left panel is devoted to the three first modes ($m=1,2,3$) with $n=0,1$ and $k_z=0$. Notice that the $u_z$ is independent of the core-shell NW materials involved. The right panel presents the elongation for $n=0$ and $k_z\neq0$ for Ge-Si and Si-Ge NWs.

\subsection{Polarization along the radial direction}

The vector component $u_{r}^{(\hat{e}_{r})}$ is a mixture of the tree amplitudes, $u_{L}$, $u_{T1}$ and $u_{T2}$; thus, employing Eq.~(\ref{basis}) we have

\begin{multline}
u_{r}^{(\hat{e}_{r})}=U_{r}e^{in\theta }=\left( A_{T2}J_{n}^{\text{ }\prime}(\mu_{T2}r/a)+A_{T1}J_{n}(\mu_{T1}r/a)\right. \\
\left.+J_{n}^{\text{ }\prime }(\mu_{L}r/a)\right) e^{in\theta }/\sqrt{N_{r}} \text{ },
\label{Uradio}
\end{multline}
where the constants $A_{T1}$ and $A_{T2}$ are given in the Appendix. This allows us to reduce the matrix elements (\ref{M}) to
\begin{multline}
M_{\beta ,\alpha }^{(\hat{e}_{r})}=\left( \delta _{\nu ^{\prime },\nu + n+2}
\left[ -\langle \nu ^{\prime }|U_{r}|\nu +1\rangle +\right. \right. \\
\left. \langle \nu ^{\prime }+2|U_{r}|\nu +3\rangle \right] +\delta _{\nu^{\prime },\nu +n-2}\left[ -\langle \nu ^{\prime }+1|U_{z}|\nu \rangle \right. \\
\left. \left. +\langle \nu ^{\prime }+3|U_{r}|\nu +2\rangle \right] \right)
\delta_{k_{e}^{\prime }, k_{e}\pm k_{z}}\text{ }.
\label{Mr}
\end{multline}

Notice that even for $k_{z}=0$ the EPH $H_{\rm e-ph}^{(\hat{e}_{r})}$ present a mixture of the $L-T2$ modes. Only for $n=0$ there is a pure longitudinal oscillation along the radial direction.

\begin{figure}[tbh]
\begin{center}
\includegraphics[width=\columnwidth]{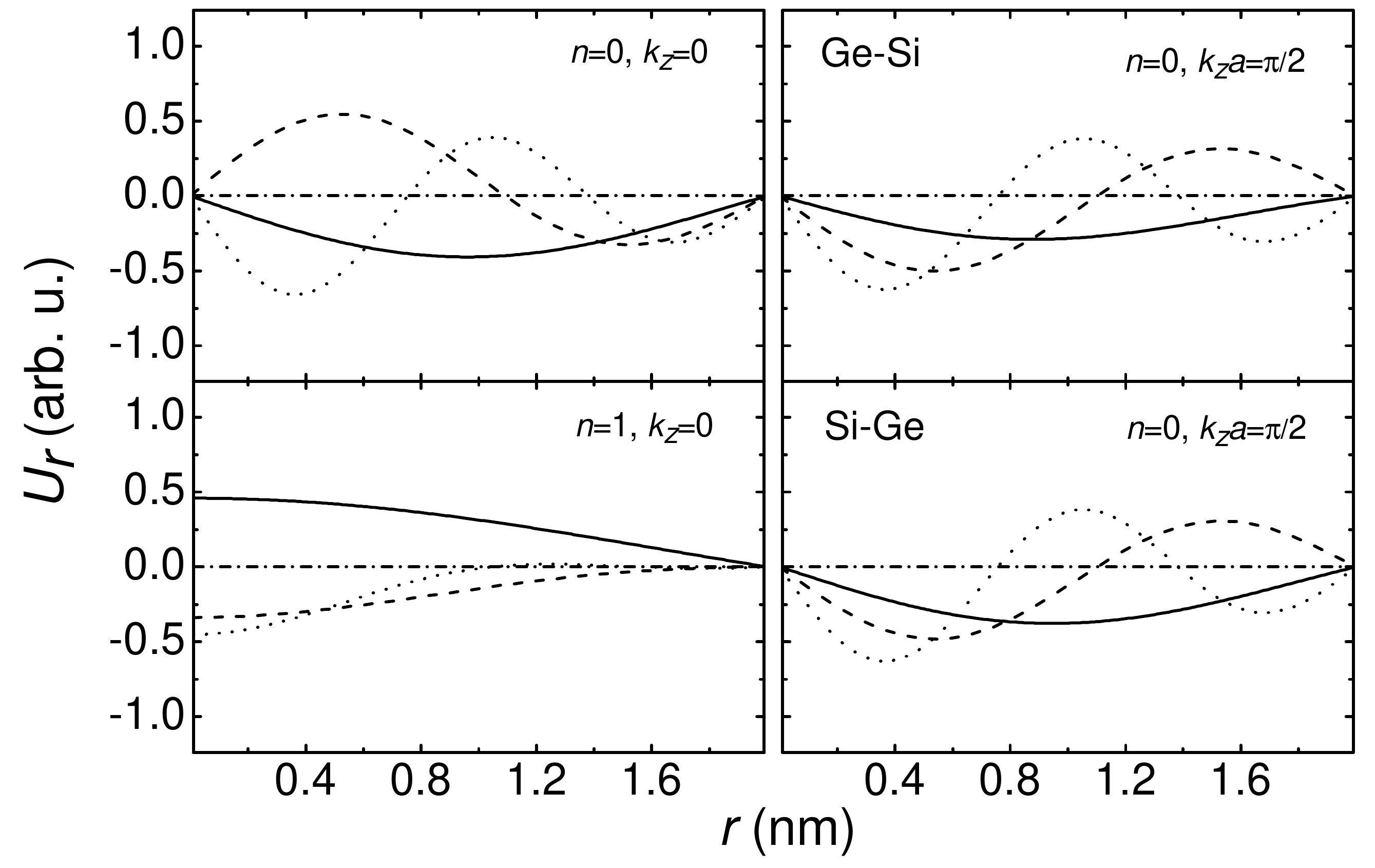}\\[0pt]
\end{center}
\caption{Same as in Fig.~(\ref{Uz}) for core phonon amplitude $U_{r}$. }
\label{Ur}
\end{figure}

\subsection{Polarization along the azimuthal direction }

From the basis given in  Eq.~(\ref{basis}) we have

\begin{multline}
u_{\theta }^{(\hat{e}_{\theta })} = U_{\theta}e^{in\theta }=\left(B_{T1}J_{n}(\mu_{T1}r/a)-J_{n}^{\text{ }\prime }(\mu_{T2}r/a)\right. \\
\left. + B_{L}J_{n}(\mu_{L}r/a)\right) e^{in\theta }/\sqrt{N_{r}}\text{ },
\label{Utheta}
\end{multline}

\noindent where the coefficients $B_{T1}$ and $B_{L}$ are reported in the Appendix.

\begin{figure}[tbh]
\begin{center}
\includegraphics[width=\columnwidth]{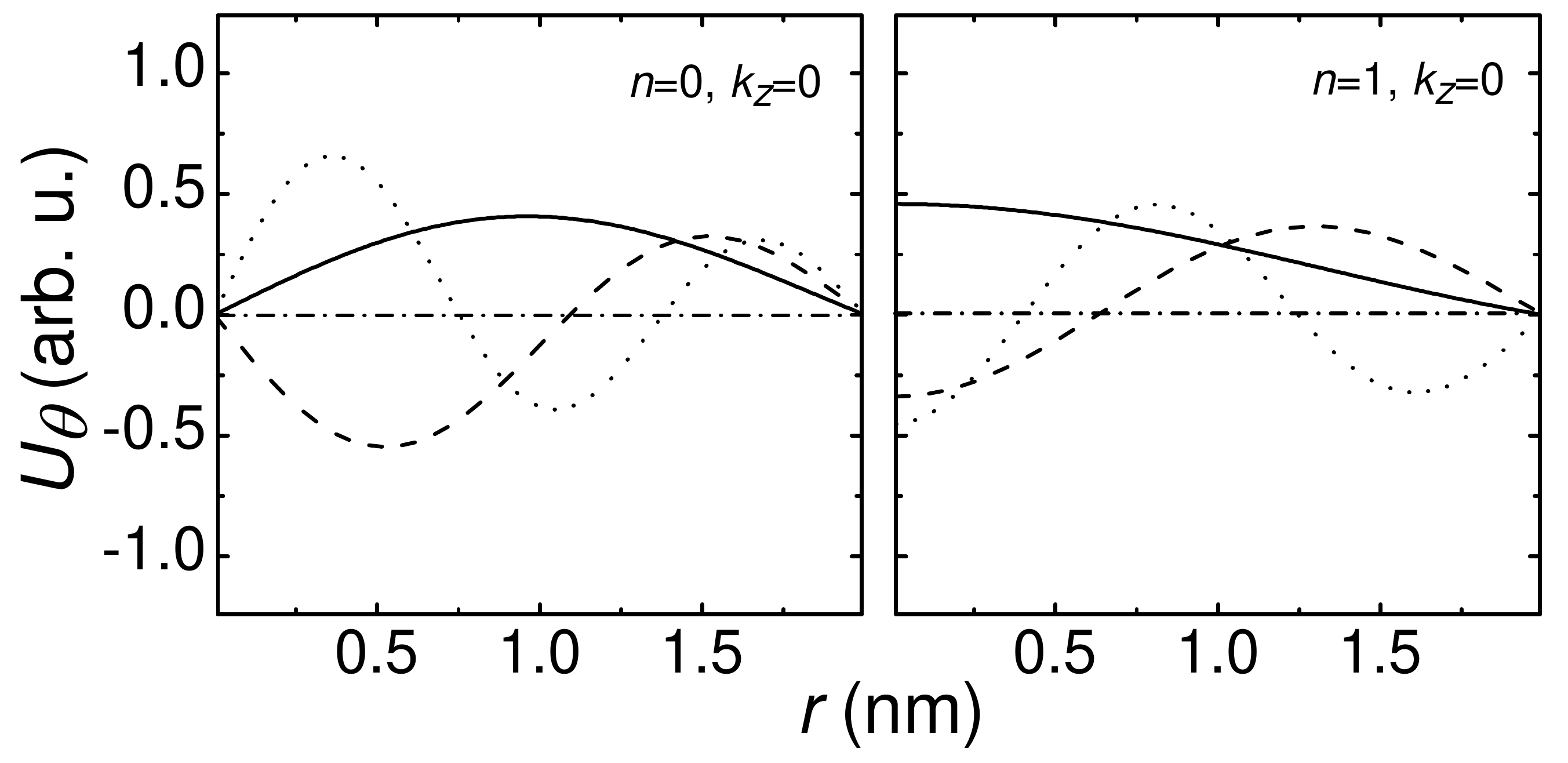}\\[0pt]
\end{center}
\caption{Core phonon amplitude $U_{\theta}$ for Ge-Si and Si-Ge core-shell NW and $k_{z}=0$. Left panel $n=0$; right panel $n=1$. In the calculation $a=2$ nm and $b=4$ nm.}
\label{Uthe}
\end{figure}

With this latter expression, the scattering matrix element with a deformation potential $D_{(\hat{e}_{\theta })}$ becomes

\begin{multline}
M_{\beta ,\alpha }^{(\hat{e}_{\theta })}=\left( \delta _{\nu ^{\prime },\nu
+n+2}\left[ -\langle \nu ^{\prime }|U_{\theta }|\nu +1\rangle -\right. \right. \\
\left. \langle \nu ^{\prime }+2|U_{\theta }|\nu +3\rangle \right] + \delta_{\nu ^{\prime },\nu +n-2}\left[ \langle \nu ^{\prime }+1|U_{\theta }|\nu
\rangle \right. \\
\left. \left. +\langle \nu ^{\prime }+3|U_{\theta }|\nu +2\rangle \right]
\right) \delta _{k_{e}^{\prime },k_{e}\pm k_{z}}\text{ }.
\label{Mtheta}
\end{multline}

The dependence on $r$ of the phonon elongations $U_r$ and $U_\theta$ which  appear in $H_{\rm e-ph}^{(\hat{e}_{r})}$ and  $H_{\rm e-ph}^{(\hat{e}_{\theta})}$ are shown in Figs.~\ref{Ur} and \ref{Uthe} respectively. For both, Si-Ge and Ge-Si NWs we take  $n=0, 1$, $k_za=0$, $\pi /2$  and $m=1, 2, 3$.

Notice that the deformation potential scattering amplitudes (\ref{M}) of the special reported cases given by Eqs.~(\ref{Mz}), (\ref{Mr}) and (\ref{Mtheta}) take into account the phonon symmetries of Ge-Si and Si-Ge NWs and the corresponding strain effects. All information of shell structure is carried out in the phonon symmetry and frequency calculations $\omega_{m,n}=\omega_{m,n}(a,\lambda_{i}(\gamma ))$.

\section{\label{Conclusions}Conclusions}

In this work we presented complete treatment of the non-polar optical phonons and the electron-phonon deformation potential interaction in core-shell cylindrical nanowires. The vector phonon displacement field $\vec{U}$ is derived by solving a system of coupled differential equations providing a general basis for the solutions of the problem. It is found that the modes shows mixed torsional, axial and radial characters, depending on the physical conditions involved. Thus, in general, the $H_{\rm e-ph}$ cannot
be decoupled into pure transversal or longitudinal motion and depending on the phonon propagation direction with respect nanowire axis.

So long as the phonon amplitude $U$ has typical dimensions scaling as the core size, $a$, and deformation coupling constant in Eqs. (\ref{Matrix_Dr})-(\ref{Matrix_Dz}) is proportional to $a^{-\frac{3}{2}}$, the magnitude of deformation potential Hamiltonian is proportional to $1/\sqrt{a}$. Hence, the electron-phonon interaction increases as the core radius decreases and the effects of the mechanical boundary conditions become important. Similar result have been reported and observed experimentally in spherical quantum dots.~\cite{PhysRevB.78.081304}

For the case of dressed non-polar Si-Ge-based nanowires with complete confinement, the shell has a role on the frequency shift of the core optical modes through the strain.  The employed basis (\ref{basis}) for the solutions of the problem, allows to study the influence of the longitudinal and transversal mixtures on $H_{\rm e-ph}$ as function of the confinement and wavevector $k_{z}$. Also, we give explicit analytical expressions of the $H_{\rm e-ph}$ for the cases of torsional, axial and radial phonon propagations. Moreover, electronic transitions in the valence are assisted by phonons if the angular momentum quantum numbers for the involved hole states fulfill the selection rule delta $\Delta\nu = n + 2$ related to the emission or absorption of one confined phonon.

Our model allows for the study of optical phonon deformation potential as a function of the structural parameters, which contains crucial information for the characterization of core-shell nanowires.

\appendix
\section{Phonon amplitudes}

The coefficients for the phonon elongation $Ur$ in Eq.~(\ref{Uradio}) are given by

\begin{multline}
A_{T1}=-\left( \frac{\widetilde{k}_{z}^{2}
}{\mu_{T1}\mu_{L}}\frac{J_{n}(\mu_{L})J_{n}^{\text{ }\prime \text{ }}(\mu_{T2})}{J^2_{n}(\mu_{T1})}+\frac{J_{n}^{\text{ }\prime \text{ }}(\mu_{L})}{J_{n}(\mu_{T1})}\right) \text{ },
\end{multline}

and

\begin{equation}
A_{T2}=\frac{\widetilde{k}_{z}^{2}}{\mu_{T1}\mu_{L}}\frac{
J_{n}(\mu_{L})}{J_{n}(\mu_{T1})}~.
\end{equation}

For the amplitude $U_{\theta }$ in Eq.~(\ref{Utheta}) we obtain

\begin{eqnarray}
B_{T1}=\frac{\widetilde{k}_{z}^{2}}{\widetilde{k}_{z}^{2}+\mu_{T1}^{2}}\frac{J_{n}^{\text{ }\prime }(\mu_{T2})}{J_{n}(\mu_{T1})}~, \\
B_{L}=\frac{\mu_{T1}^{2}}{\widetilde{k}_{z}^{2}+\mu_{T1}^{2}}\frac{J_{n}^{\text{ }\prime }(\mu_{T2})}{J_{n}(\mu_{L})}~.
\end{eqnarray}

\acknowledgments
This work was partially supported by Spanish MINECO through Grant FIS2012-33521. D. S-P, C. T-G and G. E. M acknowledge support from the
Brazilian Agencies FAPESP and CNPq. R.P.-A. acknowledges CONACyT (M\'exico) support through grant 208108 and hospitality at ICMM-CSIC, Madrid, Spain.

\bibliography{Bib-Otros,Bib-NanoWires-SiGe,Bib-NanoWires-Ge,Bib-NanoWires-Si}

\begin{thebibliography}{44}%
\makeatletter
\providecommand \@ifxundefined [1]{%
 \@ifx{#1\undefined}
}%
\providecommand \@ifnum [1]{%
 \ifnum #1\expandafter \@firstoftwo
 \else \expandafter \@secondoftwo
 \fi
}%
\providecommand \@ifx [1]{%
 \ifx #1\expandafter \@firstoftwo
 \else \expandafter \@secondoftwo
 \fi
}%
\providecommand \natexlab [1]{#1}%
\providecommand \enquote  [1]{``#1''}%
\providecommand \bibnamefont  [1]{#1}%
\providecommand \bibfnamefont [1]{#1}%
\providecommand \citenamefont [1]{#1}%
\providecommand \href@noop [0]{\@secondoftwo}%
\providecommand \href [0]{\begingroup \@sanitize@url \@href}%
\providecommand \@href[1]{\@@startlink{#1}\@@href}%
\providecommand \@@href[1]{\endgroup#1\@@endlink}%
\providecommand \@sanitize@url [0]{\catcode `\\12\catcode `\$12\catcode
  `\&12\catcode `\#12\catcode `\^12\catcode `\_12\catcode `\%12\relax}%
\providecommand \@@startlink[1]{}%
\providecommand \@@endlink[0]{}%
\providecommand \url  [0]{\begingroup\@sanitize@url \@url }%
\providecommand \@url [1]{\endgroup\@href {#1}{\urlprefix }}%
\providecommand \urlprefix  [0]{URL }%
\providecommand \Eprint [0]{\href }%
\providecommand \doibase [0]{http://dx.doi.org/}%
\providecommand \selectlanguage [0]{\@gobble}%
\providecommand \bibinfo  [0]{\@secondoftwo}%
\providecommand \bibfield  [0]{\@secondoftwo}%
\providecommand \translation [1]{[#1]}%
\providecommand \BibitemOpen [0]{}%
\providecommand \bibitemStop [0]{}%
\providecommand \bibitemNoStop [0]{.\EOS\space}%
\providecommand \EOS [0]{\spacefactor3000\relax}%
\providecommand \BibitemShut  [1]{\csname bibitem#1\endcsname}%
\let\auto@bib@innerbib\@empty
\bibitem [{\citenamefont {Lu}\ and\ \citenamefont
  {Lieber}(2006)}]{WuLieber2006}%
  \BibitemOpen
  \bibfield  {author} {\bibinfo {author} {\bibfnamefont {W.}~\bibnamefont
  {Lu}}\ and\ \bibinfo {author} {\bibfnamefont {C.~M.}\ \bibnamefont
  {Lieber}},\ }\href {http://stacks.iop.org/0022-3727/39/i=21/a=R01} {\bibfield
   {journal} {\bibinfo  {journal} {Journal of Physics D: Applied Physics}\
  }\textbf {\bibinfo {volume} {39}},\ \bibinfo {pages} {R387} (\bibinfo {year}
  {2006})}\BibitemShut {NoStop}%
\bibitem [{\citenamefont {Hayden}\ \emph {et~al.}(2008)\citenamefont {Hayden},
  \citenamefont {Agarwal},\ and\ \citenamefont {Lu}}]{Hayden2008}%
  \BibitemOpen
  \bibfield  {author} {\bibinfo {author} {\bibfnamefont {O.}~\bibnamefont
  {Hayden}}, \bibinfo {author} {\bibfnamefont {R.}~\bibnamefont {Agarwal}}, \
  and\ \bibinfo {author} {\bibfnamefont {W.}~\bibnamefont {Lu}},\ }\href
  {\doibase http://dx.doi.org/10.1016/S1748-0132(08)70061-6} {\bibfield
  {journal} {\bibinfo  {journal} {Nano Today}\ }\textbf {\bibinfo {volume}
  {3}},\ \bibinfo {pages} {12 } (\bibinfo {year} {2008})}\BibitemShut {NoStop}%
\bibitem [{\citenamefont {Dasgupta}\ \emph {et~al.}(2014)\citenamefont
  {Dasgupta}, \citenamefont {Sun}, \citenamefont {Liu}, \citenamefont
  {Brittman}, \citenamefont {Andrews}, \citenamefont {Lim}, \citenamefont
  {Gao}, \citenamefont {Yan},\ and\ \citenamefont {Yang}}]{AdvmatDasgupta2014}%
  \BibitemOpen
  \bibfield  {author} {\bibinfo {author} {\bibfnamefont {N.~P.}\ \bibnamefont
  {Dasgupta}}, \bibinfo {author} {\bibfnamefont {J.}~\bibnamefont {Sun}},
  \bibinfo {author} {\bibfnamefont {C.}~\bibnamefont {Liu}}, \bibinfo {author}
  {\bibfnamefont {S.}~\bibnamefont {Brittman}}, \bibinfo {author}
  {\bibfnamefont {S.~C.}\ \bibnamefont {Andrews}}, \bibinfo {author}
  {\bibfnamefont {J.}~\bibnamefont {Lim}}, \bibinfo {author} {\bibfnamefont
  {H.}~\bibnamefont {Gao}}, \bibinfo {author} {\bibfnamefont {R.}~\bibnamefont
  {Yan}}, \ and\ \bibinfo {author} {\bibfnamefont {P.}~\bibnamefont {Yang}},\
  }\href {\doibase 10.1002/adma.201305929} {\bibfield  {journal} {\bibinfo
  {journal} {Advanced Materials}\ }\textbf {\bibinfo {volume} {26}},\ \bibinfo
  {pages} {2137} (\bibinfo {year} {2014})}\BibitemShut {NoStop}%
\bibitem [{\citenamefont {Lauhon}\ \emph {et~al.}(2002)\citenamefont {Lauhon},
  \citenamefont {Gudiksen}, \citenamefont {Wang},\ and\ \citenamefont
  {Lieber}}]{Lauhon2002}%
  \BibitemOpen
  \bibfield  {author} {\bibinfo {author} {\bibfnamefont {L.~J.}\ \bibnamefont
  {Lauhon}}, \bibinfo {author} {\bibfnamefont {M.~S.}\ \bibnamefont
  {Gudiksen}}, \bibinfo {author} {\bibfnamefont {D.}~\bibnamefont {Wang}}, \
  and\ \bibinfo {author} {\bibfnamefont {C.~M.}\ \bibnamefont {Lieber}},\
  }\href@noop {} {\bibfield  {journal} {\bibinfo  {journal} {Nature}\ }\textbf
  {\bibinfo {volume} {420}},\ \bibinfo {pages} {57} (\bibinfo {year}
  {2002})}\BibitemShut {NoStop}%
\bibitem [{\citenamefont {Singh}\ \emph {et~al.}(2012)\citenamefont {Singh},
  \citenamefont {Poweleit}, \citenamefont {Dailey}, \citenamefont {Drucker},\
  and\ \citenamefont {Men\'endez}}]{SinghSST2012}%
  \BibitemOpen
  \bibfield  {author} {\bibinfo {author} {\bibfnamefont {R.}~\bibnamefont
  {Singh}}, \bibinfo {author} {\bibfnamefont {C.~D.}\ \bibnamefont {Poweleit}},
  \bibinfo {author} {\bibfnamefont {E.}~\bibnamefont {Dailey}}, \bibinfo
  {author} {\bibfnamefont {J.}~\bibnamefont {Drucker}}, \ and\ \bibinfo
  {author} {\bibfnamefont {J.}~\bibnamefont {Men\'endez}},\ }\href
  {http://stacks.iop.org/0268-1242/27/i=8/a=085008} {\bibfield  {journal}
  {\bibinfo  {journal} {Semiconductor Science and Technology}\ }\textbf
  {\bibinfo {volume} {27}},\ \bibinfo {pages} {085008} (\bibinfo {year}
  {2012})}\BibitemShut {NoStop}%
\bibitem [{\citenamefont {Dillen}\ \emph {et~al.}(2012)\citenamefont {Dillen},
  \citenamefont {Varahramyan}, \citenamefont {Corbet},\ and\ \citenamefont
  {Tutuc}}]{PhysRevB.86.045311}%
  \BibitemOpen
  \bibfield  {author} {\bibinfo {author} {\bibfnamefont {D.~C.}\ \bibnamefont
  {Dillen}}, \bibinfo {author} {\bibfnamefont {K.~M.}\ \bibnamefont
  {Varahramyan}}, \bibinfo {author} {\bibfnamefont {C.~M.}\ \bibnamefont
  {Corbet}}, \ and\ \bibinfo {author} {\bibfnamefont {E.}~\bibnamefont
  {Tutuc}},\ }\href {\doibase 10.1103/PhysRevB.86.045311} {\bibfield  {journal}
  {\bibinfo  {journal} {Phys. Rev. B}\ }\textbf {\bibinfo {volume} {86}},\
  \bibinfo {pages} {045311} (\bibinfo {year} {2012})}\BibitemShut {NoStop}%
\bibitem [{\citenamefont {Kallel}\ \emph {et~al.}(2012)\citenamefont {Kallel},
  \citenamefont {Arbouet}, \citenamefont {BenAssayag}, \citenamefont
  {Chehaidar}, \citenamefont {Poti\'e}, \citenamefont {Salem}, \citenamefont
  {Baron},\ and\ \citenamefont {Paillard}}]{PhysRevB.86.085318}%
  \BibitemOpen
  \bibfield  {author} {\bibinfo {author} {\bibfnamefont {H.}~\bibnamefont
  {Kallel}}, \bibinfo {author} {\bibfnamefont {A.}~\bibnamefont {Arbouet}},
  \bibinfo {author} {\bibfnamefont {G.}~\bibnamefont {BenAssayag}}, \bibinfo
  {author} {\bibfnamefont {A.}~\bibnamefont {Chehaidar}}, \bibinfo {author}
  {\bibfnamefont {A.}~\bibnamefont {Poti\'e}}, \bibinfo {author} {\bibfnamefont
  {B.}~\bibnamefont {Salem}}, \bibinfo {author} {\bibfnamefont
  {T.}~\bibnamefont {Baron}}, \ and\ \bibinfo {author} {\bibfnamefont
  {V.}~\bibnamefont {Paillard}},\ }\href {\doibase 10.1103/PhysRevB.86.085318}
  {\bibfield  {journal} {\bibinfo  {journal} {Phys. Rev. B}\ }\textbf {\bibinfo
  {volume} {86}},\ \bibinfo {pages} {085318} (\bibinfo {year}
  {2012})}\BibitemShut {NoStop}%
\bibitem [{\citenamefont {Singh}\ \emph {et~al.}(2011)\citenamefont {Singh},
  \citenamefont {Dailey}, \citenamefont {Drucker},\ and\ \citenamefont
  {Men\'endez}}]{singh:124305}%
  \BibitemOpen
  \bibfield  {author} {\bibinfo {author} {\bibfnamefont {R.}~\bibnamefont
  {Singh}}, \bibinfo {author} {\bibfnamefont {E.~J.}\ \bibnamefont {Dailey}},
  \bibinfo {author} {\bibfnamefont {J.}~\bibnamefont {Drucker}}, \ and\
  \bibinfo {author} {\bibfnamefont {J.}~\bibnamefont {Men\'endez}},\ }\href
  {\doibase 10.1063/1.3667125} {\bibfield  {journal} {\bibinfo  {journal}
  {Journal of Applied Physics}\ }\textbf {\bibinfo {volume} {110}},\ \bibinfo
  {eid} {124305} (\bibinfo {year} {2011})}\BibitemShut {NoStop}%
\bibitem [{\citenamefont {Mart\'{\i}nez-Guti\'errez}\ and\ \citenamefont
  {Velasco}(2013)}]{MartinezGutierrez201386}%
  \BibitemOpen
  \bibfield  {author} {\bibinfo {author} {\bibfnamefont {D.}~\bibnamefont
  {Mart\'{\i}nez-Guti\'errez}}\ and\ \bibinfo {author} {\bibfnamefont
  {V.}~\bibnamefont {Velasco}},\ }\href {\doibase
  http://dx.doi.org/10.1016/j.physe.2013.05.025} {\bibfield  {journal}
  {\bibinfo  {journal} {Physica E: Low-dimensional Systems and Nanostructures}\
  }\textbf {\bibinfo {volume} {54}},\ \bibinfo {pages} {86 } (\bibinfo {year}
  {2013})}\BibitemShut {NoStop}%
\bibitem [{\citenamefont {Hu}\ \emph {et~al.}(2011)\citenamefont {Hu},
  \citenamefont {Zhang}, \citenamefont {Giapis},\ and\ \citenamefont
  {Poulikakos}}]{PhysRevB.84.085442}%
  \BibitemOpen
  \bibfield  {author} {\bibinfo {author} {\bibfnamefont {M.}~\bibnamefont
  {Hu}}, \bibinfo {author} {\bibfnamefont {X.}~\bibnamefont {Zhang}}, \bibinfo
  {author} {\bibfnamefont {K.~P.}\ \bibnamefont {Giapis}}, \ and\ \bibinfo
  {author} {\bibfnamefont {D.}~\bibnamefont {Poulikakos}},\ }\href {\doibase
  10.1103/PhysRevB.84.085442} {\bibfield  {journal} {\bibinfo  {journal} {Phys.
  Rev. B}\ }\textbf {\bibinfo {volume} {84}},\ \bibinfo {pages} {085442}
  (\bibinfo {year} {2011})}\BibitemShut {NoStop}%
\bibitem [{\citenamefont {Liu}\ \emph {et~al.}(2008)\citenamefont {Liu},
  \citenamefont {Hu},\ and\ \citenamefont {Pan}}]{Liu20083042}%
  \BibitemOpen
  \bibfield  {author} {\bibinfo {author} {\bibfnamefont {X.}~\bibnamefont
  {Liu}}, \bibinfo {author} {\bibfnamefont {J.}~\bibnamefont {Hu}}, \ and\
  \bibinfo {author} {\bibfnamefont {B.}~\bibnamefont {Pan}},\ }\href {\doibase
  http://dx.doi.org/10.1016/j.physe.2008.03.011} {\bibfield  {journal}
  {\bibinfo  {journal} {Physica E: Low-dimensional Systems and Nanostructures}\
  }\textbf {\bibinfo {volume} {40}},\ \bibinfo {pages} {3042 } (\bibinfo {year}
  {2008})}\BibitemShut {NoStop}%
\bibitem [{\citenamefont {Pek\"oz}\ and\ \citenamefont
  {Raty}(2009)}]{PhysRevB.80.155432}%
  \BibitemOpen
  \bibfield  {author} {\bibinfo {author} {\bibfnamefont {R.}~\bibnamefont
  {Pek\"oz}}\ and\ \bibinfo {author} {\bibfnamefont {J.-Y.}\ \bibnamefont
  {Raty}},\ }\href {\doibase 10.1103/PhysRevB.80.155432} {\bibfield  {journal}
  {\bibinfo  {journal} {Phys. Rev. B}\ }\textbf {\bibinfo {volume} {80}},\
  \bibinfo {pages} {155432} (\bibinfo {year} {2009})}\BibitemShut {NoStop}%
\bibitem [{\citenamefont {Musin}\ and\ \citenamefont
  {Wang}(2005)}]{PhysRevB.71.155318}%
  \BibitemOpen
  \bibfield  {author} {\bibinfo {author} {\bibfnamefont {R.~N.}\ \bibnamefont
  {Musin}}\ and\ \bibinfo {author} {\bibfnamefont {X.-Q.}\ \bibnamefont
  {Wang}},\ }\href {\doibase 10.1103/PhysRevB.71.155318} {\bibfield  {journal}
  {\bibinfo  {journal} {Phys. Rev. B}\ }\textbf {\bibinfo {volume} {71}},\
  \bibinfo {pages} {155318} (\bibinfo {year} {2005})}\BibitemShut {NoStop}%
\bibitem [{\citenamefont {Chan}\ \emph {et~al.}(2010)\citenamefont {Chan},
  \citenamefont {Reed}, \citenamefont {Donadio}, \citenamefont {Mueller},
  \citenamefont {Meng}, \citenamefont {Galli},\ and\ \citenamefont
  {Ceder}}]{PhysRevB.81.174303}%
  \BibitemOpen
  \bibfield  {author} {\bibinfo {author} {\bibfnamefont {M.~K.~Y.}\
  \bibnamefont {Chan}}, \bibinfo {author} {\bibfnamefont {J.}~\bibnamefont
  {Reed}}, \bibinfo {author} {\bibfnamefont {D.}~\bibnamefont {Donadio}},
  \bibinfo {author} {\bibfnamefont {T.}~\bibnamefont {Mueller}}, \bibinfo
  {author} {\bibfnamefont {Y.~S.}\ \bibnamefont {Meng}}, \bibinfo {author}
  {\bibfnamefont {G.}~\bibnamefont {Galli}}, \ and\ \bibinfo {author}
  {\bibfnamefont {G.}~\bibnamefont {Ceder}},\ }\href {\doibase
  10.1103/PhysRevB.81.174303} {\bibfield  {journal} {\bibinfo  {journal} {Phys.
  Rev. B}\ }\textbf {\bibinfo {volume} {81}},\ \bibinfo {pages} {174303}
  (\bibinfo {year} {2010})}\BibitemShut {NoStop}%
\bibitem [{\citenamefont {Trammell}\ \emph {et~al.}(2008)\citenamefont
  {Trammell}, \citenamefont {Zhang}, \citenamefont {Li}, \citenamefont {Chen},\
  and\ \citenamefont {Dickey}}]{Trammell2008}%
  \BibitemOpen
  \bibfield  {author} {\bibinfo {author} {\bibfnamefont {T.~E.}\ \bibnamefont
  {Trammell}}, \bibinfo {author} {\bibfnamefont {X.}~\bibnamefont {Zhang}},
  \bibinfo {author} {\bibfnamefont {Y.}~\bibnamefont {Li}}, \bibinfo {author}
  {\bibfnamefont {L.-Q.}\ \bibnamefont {Chen}}, \ and\ \bibinfo {author}
  {\bibfnamefont {E.~C.}\ \bibnamefont {Dickey}},\ }\href {\doibase
  http://dx.doi.org/10.1016/j.jcrysgro.2008.02.037} {\bibfield  {journal}
  {\bibinfo  {journal} {Journal of Crystal Growth}\ }\textbf {\bibinfo {volume}
  {310}},\ \bibinfo {pages} {3084 } (\bibinfo {year} {2008})}\BibitemShut
  {NoStop}%
\bibitem [{\citenamefont {Goldthorpe}\ \emph {et~al.}(2008)\citenamefont
  {Goldthorpe}, \citenamefont {Marshall},\ and\ \citenamefont
  {McIntyre}}]{Goldthorpe2008}%
  \BibitemOpen
  \bibfield  {author} {\bibinfo {author} {\bibfnamefont {I.~A.}\ \bibnamefont
  {Goldthorpe}}, \bibinfo {author} {\bibfnamefont {A.~F.}\ \bibnamefont
  {Marshall}}, \ and\ \bibinfo {author} {\bibfnamefont {P.~C.}\ \bibnamefont
  {McIntyre}},\ }\href {\doibase 10.1021/nl802408y} {\bibfield  {journal}
  {\bibinfo  {journal} {Nano Letters}\ }\textbf {\bibinfo {volume} {8}},\
  \bibinfo {pages} {4081} (\bibinfo {year} {2008})}\BibitemShut {NoStop}%
\bibitem [{\citenamefont {Kloeffel}\ \emph {et~al.}(2014)\citenamefont
  {Kloeffel}, \citenamefont {Trif},\ and\ \citenamefont {Loss}}]{Loss2014}%
  \BibitemOpen
  \bibfield  {author} {\bibinfo {author} {\bibfnamefont {C.}~\bibnamefont
  {Kloeffel}}, \bibinfo {author} {\bibfnamefont {M.}~\bibnamefont {Trif}}, \
  and\ \bibinfo {author} {\bibfnamefont {D.}~\bibnamefont {Loss}},\ }\href
  {\doibase 10.1103/PhysRevB.90.115419} {\bibfield  {journal} {\bibinfo
  {journal} {Phys. Rev. B}\ }\textbf {\bibinfo {volume} {90}},\ \bibinfo
  {pages} {115419} (\bibinfo {year} {2014})}\BibitemShut {NoStop}%
\bibitem [{\citenamefont {Maier}\ and\ \citenamefont {Loss}(2012)}]{Loss2012}%
  \BibitemOpen
  \bibfield  {author} {\bibinfo {author} {\bibfnamefont {F.}~\bibnamefont
  {Maier}}\ and\ \bibinfo {author} {\bibfnamefont {D.}~\bibnamefont {Loss}},\
  }\href {\doibase 10.1103/PhysRevB.85.195323} {\bibfield  {journal} {\bibinfo
  {journal} {Phys. Rev. B}\ }\textbf {\bibinfo {volume} {85}},\ \bibinfo
  {pages} {195323} (\bibinfo {year} {2012})}\BibitemShut {NoStop}%
\bibitem [{\citenamefont {Rolo}\ \emph {et~al.}(2008)\citenamefont {Rolo},
  \citenamefont {Vasilevskiy}, \citenamefont {Hamma},\ and\ \citenamefont
  {Trallero-Giner}}]{PhysRevB.78.081304}%
  \BibitemOpen
  \bibfield  {author} {\bibinfo {author} {\bibfnamefont {A.~G.}\ \bibnamefont
  {Rolo}}, \bibinfo {author} {\bibfnamefont {M.~I.}\ \bibnamefont
  {Vasilevskiy}}, \bibinfo {author} {\bibfnamefont {M.}~\bibnamefont {Hamma}},
  \ and\ \bibinfo {author} {\bibfnamefont {C.}~\bibnamefont {Trallero-Giner}},\
  }\href {\doibase 10.1103/PhysRevB.78.081304} {\bibfield  {journal} {\bibinfo
  {journal} {Phys. Rev. B}\ }\textbf {\bibinfo {volume} {78}},\ \bibinfo
  {pages} {081304} (\bibinfo {year} {2008})}\BibitemShut {NoStop}%
\bibitem [{\citenamefont {Bir}\ and\ \citenamefont
  {Pikus}(1974)}]{bir1974symmetry}%
  \BibitemOpen
  \bibfield  {author} {\bibinfo {author} {\bibfnamefont {G.}~\bibnamefont
  {Bir}}\ and\ \bibinfo {author} {\bibfnamefont {G.}~\bibnamefont {Pikus}},\
  }\href@noop {} {\emph {\bibinfo {title} {Symmetry and Strain-induced Effects
  in Semiconductors}}},\ A Halsted press book\ (\bibinfo  {publisher} {Wiley},\
  \bibinfo {address} {New York},\ \bibinfo {year} {1974})\BibitemShut {NoStop}%
\bibitem [{\citenamefont {Men\'{e}ndez}\ \emph {et~al.}(2011)\citenamefont
  {Men\'{e}ndez}, \citenamefont {Singh},\ and\ \citenamefont
  {Drucker}}]{Menendez}%
  \BibitemOpen
  \bibfield  {author} {\bibinfo {author} {\bibfnamefont {J.}~\bibnamefont
  {Men\'{e}ndez}}, \bibinfo {author} {\bibfnamefont {R.}~\bibnamefont {Singh}},
  \ and\ \bibinfo {author} {\bibfnamefont {J.}~\bibnamefont {Drucker}},\
  }\href@noop {} {\bibfield  {journal} {\bibinfo  {journal} {Ann. Phys.
  (Berlin)}\ }\textbf {\bibinfo {volume} {523}},\ \bibinfo {pages} {145}
  (\bibinfo {year} {2011})}\BibitemShut {NoStop}%
\bibitem [{\citenamefont {Trallero-Giner}\ \emph {et~al.}(1998)\citenamefont
  {Trallero-Giner}, \citenamefont {P\'erez-\'Alvarez},\ and\ \citenamefont
  {Garc\'{\i}a-Moliner}}]{CubaLibro}%
  \BibitemOpen
  \bibfield  {author} {\bibinfo {author} {\bibfnamefont {C.}~\bibnamefont
  {Trallero-Giner}}, \bibinfo {author} {\bibfnamefont {R.}~\bibnamefont
  {P\'erez-\'Alvarez}}, \ and\ \bibinfo {author} {\bibfnamefont
  {F.}~\bibnamefont {Garc\'{\i}a-Moliner}},\ }\href@noop {} {\emph {\bibinfo
  {title} {Long wave polar modes in semiconductor heterostructures}}},\
  \bibinfo {edition} {1st}\ ed.\ (\bibinfo  {publisher} {Pergamon Elsevier
  Science},\ \bibinfo {address} {London},\ \bibinfo {year} {1998})\BibitemShut
  {NoStop}%
\bibitem [{\citenamefont {Santiago-P\'erez}\ \emph {et~al.}(2012)\citenamefont
  {Santiago-P\'erez}, \citenamefont {Trallero-Giner}, \citenamefont
  {P\'erez-\'Alvarez}, \citenamefont {Chico}, \citenamefont {Baquero},\ and\
  \citenamefont {Marques}}]{santiago-perez:084322}%
  \BibitemOpen
  \bibfield  {author} {\bibinfo {author} {\bibfnamefont {D.~G.}\ \bibnamefont
  {Santiago-P\'erez}}, \bibinfo {author} {\bibfnamefont {C.}~\bibnamefont
  {Trallero-Giner}}, \bibinfo {author} {\bibfnamefont {R.}~\bibnamefont
  {P\'erez-\'Alvarez}}, \bibinfo {author} {\bibfnamefont {L.}~\bibnamefont
  {Chico}}, \bibinfo {author} {\bibfnamefont {R.}~\bibnamefont {Baquero}}, \
  and\ \bibinfo {author} {\bibfnamefont {G.~E.}\ \bibnamefont {Marques}},\
  }\href {\doibase 10.1063/1.4761975} {\bibfield  {journal} {\bibinfo
  {journal} {Journal of Applied Physics}\ }\textbf {\bibinfo {volume} {112}},\
  \bibinfo {eid} {084322} (\bibinfo {year} {2012})}\BibitemShut {NoStop}%
\bibitem [{\citenamefont {Santiago-P\'erez}\ \emph {et~al.}(2014)\citenamefont
  {Santiago-P\'erez}, \citenamefont {Trallero-Giner}, \citenamefont
  {P\'erez-\'Alvarez},\ and\ \citenamefont {Chico}}]{SantiagoPerez2014151}%
  \BibitemOpen
  \bibfield  {author} {\bibinfo {author} {\bibfnamefont {D.~G.}\ \bibnamefont
  {Santiago-P\'erez}}, \bibinfo {author} {\bibfnamefont {C.}~\bibnamefont
  {Trallero-Giner}}, \bibinfo {author} {\bibfnamefont {R.}~\bibnamefont
  {P\'erez-\'Alvarez}}, \ and\ \bibinfo {author} {\bibfnamefont
  {L.}~\bibnamefont {Chico}},\ }\href {\doibase
  http://dx.doi.org/10.1016/j.physe.2013.08.013} {\bibfield  {journal}
  {\bibinfo  {journal} {Physica E: Low-dimensional Systems and Nanostructures}\
  }\textbf {\bibinfo {volume} {56}},\ \bibinfo {pages} {151 } (\bibinfo {year}
  {2014})}\BibitemShut {NoStop}%
\bibitem [{Note1()}]{Note1}%
  \BibitemOpen
  \bibinfo {note} {Due to symmetry reasons, the contribution of the conduction
  band at the $\Gamma $-point is zero.}\BibitemShut {Stop}%
\bibitem [{\citenamefont {Cardona}(1982)}]{Cardona}%
  \BibitemOpen
  \bibfield  {author} {\bibinfo {author} {\bibfnamefont {M.}~\bibnamefont
  {Cardona}},\ }\href@noop {} {\emph {\bibinfo {title} {Light Scattering in
  Solids II}}}\ (\bibinfo  {publisher} {Springer},\ \bibinfo {address}
  {Berlin},\ \bibinfo {year} {1982})\BibitemShut {NoStop}%
\bibitem [{\citenamefont {Madelung}(1996)}]{Madel}%
  \BibitemOpen
  \bibfield  {author} {\bibinfo {author} {\bibfnamefont {O.}~\bibnamefont
  {Madelung}},\ }\href@noop {} {\emph {\bibinfo {title} {Introduction to Solid
  State Theory}}}\ (\bibinfo  {publisher} {Springer},\ \bibinfo {address}
  {Berlin},\ \bibinfo {year} {1996})\BibitemShut {NoStop}%
\bibitem [{\citenamefont {Luttinger}\ and\ \citenamefont
  {Kohn}(1955)}]{PhysRev.97.869}%
  \BibitemOpen
  \bibfield  {author} {\bibinfo {author} {\bibfnamefont {J.~M.}\ \bibnamefont
  {Luttinger}}\ and\ \bibinfo {author} {\bibfnamefont {W.}~\bibnamefont
  {Kohn}},\ }\href {\doibase 10.1103/PhysRev.97.869} {\bibfield  {journal}
  {\bibinfo  {journal} {Phys. Rev.}\ }\textbf {\bibinfo {volume} {97}},\
  \bibinfo {pages} {869} (\bibinfo {year} {1955})}\BibitemShut {NoStop}%
\bibitem [{\citenamefont {Abramowitz}\ and\ \citenamefont
  {Stegun}(1964)}]{Abramowitz}%
  \BibitemOpen
  \bibfield  {author} {\bibinfo {author} {\bibfnamefont {M.}~\bibnamefont
  {Abramowitz}}\ and\ \bibinfo {author} {\bibfnamefont {I.}~\bibnamefont
  {Stegun}},\ }\href@noop {} {\emph {\bibinfo {title} {Handbook of Mathematical
  Functions}}}\ (\bibinfo  {publisher} {U. S. Goverment Printing Office},\
  \bibinfo {address} {Whashinton, D. C},\ \bibinfo {year} {1964})\BibitemShut
  {NoStop}%
\bibitem [{\citenamefont {Comas}\ \emph {et~al.}(1993)\citenamefont {Comas},
  \citenamefont {Trallero-Giner},\ and\ \citenamefont
  {Cantarero}}]{Comas1993a}%
  \BibitemOpen
  \bibfield  {author} {\bibinfo {author} {\bibfnamefont {F.}~\bibnamefont
  {Comas}}, \bibinfo {author} {\bibfnamefont {C.}~\bibnamefont
  {Trallero-Giner}}, \ and\ \bibinfo {author} {\bibfnamefont {A.}~\bibnamefont
  {Cantarero}},\ }\href@noop {} {\bibfield  {journal} {\bibinfo  {journal}
  {Phys. Rev. B}\ }\textbf {\bibinfo {volume} {47}},\ \bibinfo {pages} {7602}
  (\bibinfo {year} {1993})}\BibitemShut {NoStop}%
\bibitem [{\citenamefont {Comas}\ \emph {et~al.}(1995)\citenamefont {Comas},
  \citenamefont {Cantarero}, \citenamefont {Trallero-Giner},\ and\
  \citenamefont {Moshinsky}}]{0953-8984-7-9-006}%
  \BibitemOpen
  \bibfield  {author} {\bibinfo {author} {\bibfnamefont {F.}~\bibnamefont
  {Comas}}, \bibinfo {author} {\bibfnamefont {A.}~\bibnamefont {Cantarero}},
  \bibinfo {author} {\bibfnamefont {C.}~\bibnamefont {Trallero-Giner}}, \ and\
  \bibinfo {author} {\bibfnamefont {M.}~\bibnamefont {Moshinsky}},\ }\href
  {http://stacks.iop.org/0953-8984/7/i=9/a=006} {\bibfield  {journal} {\bibinfo
   {journal} {Journal of Physics: Condensed Matter}\ }\textbf {\bibinfo
  {volume} {7}},\ \bibinfo {pages} {1789} (\bibinfo {year} {1995})}\BibitemShut
  {NoStop}%
\bibitem [{\citenamefont {Chico}\ and\ \citenamefont
  {P\'erez-\'Alvarez}(2004)}]{PhysRevB.69.035419}%
  \BibitemOpen
  \bibfield  {author} {\bibinfo {author} {\bibfnamefont {L.}~\bibnamefont
  {Chico}}\ and\ \bibinfo {author} {\bibfnamefont {R.}~\bibnamefont
  {P\'erez-\'Alvarez}},\ }\href {\doibase 10.1103/PhysRevB.69.035419}
  {\bibfield  {journal} {\bibinfo  {journal} {Phys. Rev. B}\ }\textbf {\bibinfo
  {volume} {69}},\ \bibinfo {pages} {035419} (\bibinfo {year}
  {2004})}\BibitemShut {NoStop}%
\bibitem [{Note2()}]{Note2}%
  \BibitemOpen
  \bibinfo {note} {This equation is straightforward derived from the
  hydrodynamic phenomenological model for cubic polar semiconductors described
  in Refs. ~\protect \rev@citealpnum {CubaLibro,PhysRevB.37.4583} considering
  that the polarization and electric field associated with vibrations are zero.
  Note that since the core and shell bulk materials are non-polar, $\omega
  _{TO}=\omega _{LO}=\omega _{0}$.}\BibitemShut {Stop}%
\bibitem [{\citenamefont {Morse}\ and\ \citenamefont {Feshbach}(1953)}]{Morse}%
  \BibitemOpen
  \bibfield  {author} {\bibinfo {author} {\bibfnamefont {P.~M.}\ \bibnamefont
  {Morse}}\ and\ \bibinfo {author} {\bibfnamefont {H.}~\bibnamefont
  {Feshbach}},\ }\href@noop {} {\emph {\bibinfo {title} {Methods of Theoretical
  Physics}}}\ (\bibinfo  {publisher} {McGraw-Hill},\ \bibinfo {address} {New
  York},\ \bibinfo {year} {1953})\BibitemShut {NoStop}%
\bibitem [{\citenamefont {Chico}\ \emph {et~al.}(2006)\citenamefont {Chico},
  \citenamefont {P\'{e}rez-\'{A}lvarez},\ and\ \citenamefont
  {Cabrillo}}]{CPC2006}%
  \BibitemOpen
  \bibfield  {author} {\bibinfo {author} {\bibfnamefont {L.}~\bibnamefont
  {Chico}}, \bibinfo {author} {\bibfnamefont {R.}~\bibnamefont
  {P\'{e}rez-\'{A}lvarez}}, \ and\ \bibinfo {author} {\bibfnamefont
  {C.}~\bibnamefont {Cabrillo}},\ }\href@noop {} {\bibfield  {journal}
  {\bibinfo  {journal} {Phys. Rev. B}\ }\textbf {\bibinfo {volume} {73}},\
  \bibinfo {pages} {075425} (\bibinfo {year} {2006})}\BibitemShut {NoStop}%
\bibitem [{\citenamefont {Madelung}(1982)}]{Madelung}%
  \BibitemOpen
  \bibfield  {author} {\bibinfo {author} {\bibfnamefont {O.}~\bibnamefont
  {Madelung}},\ }\href@noop {} {\emph {\bibinfo {title} {Physics of Group IV
  Elements and III-V Compounds, Vol. 17a}}}\ (\bibinfo  {publisher}
  {Springer-Verlag},\ \bibinfo {address} {Berlin},\ \bibinfo {year}
  {1982})\BibitemShut {NoStop}%
\bibitem [{\citenamefont {Anastassakis}\ and\ \citenamefont
  {Cardona}(1998)}]{Anastassakis}%
  \BibitemOpen
  \bibfield  {author} {\bibinfo {author} {\bibfnamefont {E.}~\bibnamefont
  {Anastassakis}}\ and\ \bibinfo {author} {\bibfnamefont {M.}~\bibnamefont
  {Cardona}},\ }\href@noop {} {\emph {\bibinfo {title} {High Pressure in
  Semiconductor Physics II}}},\ edited by\ \bibinfo {editor} {\bibfnamefont
  {T.}~\bibnamefont {Suski}}\ and\ \bibinfo {editor} {\bibfnamefont
  {W.}~\bibnamefont {Paul}}\ (\bibinfo  {publisher} {Academic Press},\ \bibinfo
  {address} {New York},\ \bibinfo {year} {1998})\BibitemShut {NoStop}%
\bibitem [{\citenamefont {Santiago~P\'erez}\ \emph {et~al.}()\citenamefont
  {Santiago~P\'erez}, \citenamefont {Trallero-Giner}, \citenamefont
  {P\'erez-\'Alvarez},\ and\ \citenamefont {Chico}}]{phononsonly}%
  \BibitemOpen
  \bibfield  {author} {\bibinfo {author} {\bibfnamefont {D.~G.}\ \bibnamefont
  {Santiago~P\'erez}}, \bibinfo {author} {\bibfnamefont {C.}~\bibnamefont
  {Trallero-Giner}}, \bibinfo {author} {\bibfnamefont {R.}~\bibnamefont
  {P\'erez-\'Alvarez}}, \ and\ \bibinfo {author} {\bibfnamefont
  {L.}~\bibnamefont {Chico}},\ }\href@noop {} {\bibinfo  {journal}
  {unpublished}\ }\BibitemShut {NoStop}%
\bibitem [{\citenamefont {Hummer}\ \emph {et~al.}(2009)\citenamefont {Hummer},
  \citenamefont {Harl},\ and\ \citenamefont {Kresse}}]{Hummer2009}%
  \BibitemOpen
\bibfield  {journal} {  }\bibfield  {author} {\bibinfo {author} {\bibfnamefont
  {K.}~\bibnamefont {Hummer}}, \bibinfo {author} {\bibfnamefont
  {J.}~\bibnamefont {Harl}}, \ and\ \bibinfo {author} {\bibfnamefont
  {G.}~\bibnamefont {Kresse}},\ }\href {\doibase 10.1103/PhysRevB.80.115205}
  {\bibfield  {journal} {\bibinfo  {journal} {Phys. Rev. B}\ }\textbf {\bibinfo
  {volume} {80}},\ \bibinfo {pages} {115205} (\bibinfo {year}
  {2009})}\BibitemShut {NoStop}%
\bibitem [{\citenamefont {Nilsson}\ and\ \citenamefont
  {Nelin}(1972)}]{Nilsson1972}%
  \BibitemOpen
  \bibfield  {author} {\bibinfo {author} {\bibfnamefont {G.}~\bibnamefont
  {Nilsson}}\ and\ \bibinfo {author} {\bibfnamefont {G.}~\bibnamefont
  {Nelin}},\ }\href {\doibase 10.1103/PhysRevB.6.3777} {\bibfield  {journal}
  {\bibinfo  {journal} {Phys. Rev. B}\ }\textbf {\bibinfo {volume} {6}},\
  \bibinfo {pages} {3777} (\bibinfo {year} {1972})}\BibitemShut {NoStop}%
\bibitem [{\citenamefont {Kulda}\ \emph {et~al.}(1994)\citenamefont {Kulda},
  \citenamefont {Strauch}, \citenamefont {Pavone},\ and\ \citenamefont
  {Ishii}}]{Kulda1994}%
  \BibitemOpen
  \bibfield  {author} {\bibinfo {author} {\bibfnamefont {J.}~\bibnamefont
  {Kulda}}, \bibinfo {author} {\bibfnamefont {D.}~\bibnamefont {Strauch}},
  \bibinfo {author} {\bibfnamefont {P.}~\bibnamefont {Pavone}}, \ and\ \bibinfo
  {author} {\bibfnamefont {Y.}~\bibnamefont {Ishii}},\ }\href {\doibase
  10.1103/PhysRevB.50.13347} {\bibfield  {journal} {\bibinfo  {journal} {Phys.
  Rev. B}\ }\textbf {\bibinfo {volume} {50}},\ \bibinfo {pages} {13347}
  (\bibinfo {year} {1994})}\BibitemShut {NoStop}%
\bibitem [{\citenamefont {Nilsson}\ and\ \citenamefont
  {Nelin}(1971)}]{Nilsson1971}%
  \BibitemOpen
  \bibfield  {author} {\bibinfo {author} {\bibfnamefont {G.}~\bibnamefont
  {Nilsson}}\ and\ \bibinfo {author} {\bibfnamefont {G.}~\bibnamefont
  {Nelin}},\ }\href {\doibase 10.1103/PhysRevB.3.364} {\bibfield  {journal}
  {\bibinfo  {journal} {Phys. Rev. B}\ }\textbf {\bibinfo {volume} {3}},\
  \bibinfo {pages} {364} (\bibinfo {year} {1971})}\BibitemShut {NoStop}%
\bibitem [{\citenamefont {Adachi}(2005)}]{Adachi}%
  \BibitemOpen
  \bibfield  {author} {\bibinfo {author} {\bibfnamefont {S.}~\bibnamefont
  {Adachi}},\ }\href@noop {} {\emph {\bibinfo {title} {Properties of Group-IV,
  III-V and II-VI Semiconductors}}}\ (\bibinfo  {publisher} {John Wiley and
  Sons},\ \bibinfo {address} {Chichester},\ \bibinfo {year} {2005})\BibitemShut
  {NoStop}%
\bibitem [{\citenamefont {Giner}\ and\ \citenamefont
  {Comas}(1988)}]{PhysRevB.37.4583}%
  \BibitemOpen
  \bibfield  {author} {\bibinfo {author} {\bibfnamefont {C.~T.}\ \bibnamefont
  {Giner}}\ and\ \bibinfo {author} {\bibfnamefont {F.}~\bibnamefont {Comas}},\
  }\href {\doibase 10.1103/PhysRevB.37.4583} {\bibfield  {journal} {\bibinfo
  {journal} {Phys. Rev. B}\ }\textbf {\bibinfo {volume} {37}},\ \bibinfo
  {pages} {4583} (\bibinfo {year} {1988})}\BibitemShut {NoStop}%
\end{thebibliography}%

\end{document}